\documentclass[twocolumn,aps,pra,groupedaddress]{revtex4}
\usepackage{amsthm}
\usepackage{amsfonts}
\usepackage{color}
\usepackage{graphicx}
\usepackage{epsfig}
\usepackage{amssymb}
\usepackage{amsmath}
\usepackage{hyperref}
\usepackage{latexsym}
\usepackage{braket}
\usepackage{mathtools}
\usepackage{mathrsfs}
\usepackage{ulem} 

\usepackage{xcolor}

\def\overlinelz{\overline}

\def\>{\rangle}
\def\<{\langle}
\def\labell#1{\label{#1}}

\begin{document}

\title{Geometric Event-Based
  Quantum Mechanics}
  
\author{{{  Vittorio Giovannetti$^1$, Seth
  Lloyd$^{2,3}$, Lorenzo Maccone$^{2,4}$ }}} \affiliation{
  \vbox{1.~NEST, Scuola
    Normale Superiore and Istituto Nanoscienze-CNR, p.za Cavalieri 7,
    Pisa, Italy} 
    \vbox{2.~Turing Quantum, 202B Plymouth Street Brooklyn, New York
  11201, USA}\\
  \vbox{3.~MIT -- RLE and Dept. of Mech. Eng., 77
    Massachusetts Av., Cambridge, MA 02139, USA}\\
  \vbox{4.~Dip.~Fisica
    and INFN Sez.\ Pavia, University of Pavia, via Bassi 6, I-27100
    Pavia, Italy} \\
}

\begin{abstract}
  We propose a special relativistic framework for quantum mechanics.
  It is based on introducing a Hilbert space for events. Events are
  taken as primitive notions (as customary in relativity), whereas
  quantum systems (e.g.~fields and particles) are emergent in the form
  of joint probability amplitudes for position {\it and} time of
  events.  Textbook relativistic quantum mechanics and quantum field
  theory can be recovered by dividing the event Hilbert spaces into
  space and time (a foliation) and then conditioning the event states
  onto the time part. Our theory satisfies the full Poincar\'e
  symmetry as a `geometric' unitary transformation, and possesses
  observables for space (location of an event) and time (position in
  time of an event).
\end{abstract}
\pacs{}
\maketitle

Quantum Mechanics (QM) relies on time-conditioned quantities:
observables conditioned on time in the Heisenberg picture
(e.g.~$X({t})$ is the position operator at time $t$) or states
conditioned on time in the Schr\"odinger one (e.g.~$|\psi({t})\>$ is
the state at time $t$). As such, it is inherently incompatible with
the Poincar\'e symmetry of special relativity. Indeed quantum
mechanics can be formulated in a relativistic covariant fashion only
in very specific circumstances, such as Quantum Field Theory (QFT)
when using Heisenberg picture operators acting on a relativistic
invariant state such as the field vacuum.  In this paper we depart
from this approach and introduce a quantum mechanical theory of
events, the Geometric Event-Based Quantum Mechanics (GEB),
which is based on unconditioned spacetime quantities and hence
intrinsically covariant.  In QFT one starts from the dynamical
equation of motion (either in the Hamiltonian formulation or from a
Lagrangian \cite{mandl}) and quantizes the dynamical solutions
imposing equal time commutation relations. We take the opposite track:
we start by defining an (unconditioned) purely kinematic Hilbert space
${\cal H}_{\bf E}$ which is well suited to account for the symmetries
of a relativistic theory \cite{diaz,diaz1,liebrich}.  A formal
correspondence with QM and QFT is then established by showing that the
quantum evolutions defined by these theories can be identified as
special subset ${\cal H}_{\bf QM}$ of the {\it distributions} of
${\cal H}_{\bf E}$ which is determined not via dynamical equations,
but through purely geometrical
constraints~\cite{diracforms,dirachamiltoniangr}.

The Hilbert space ${\cal H}_{\bf E}$ of GEB, rather than {\it
  systems}, can be thought to describe {\it events}~\cite{lorenzo}.
In our approach the {\it event} is taken as a {\it primitive} notion,
i.e.~not something that is derived from a pre-existing notion of (say)
a pre-existing particle that has been detected (as happens in QM).
Indeed, in GEB the detection of particle at a particular location in
space and time by an inertial measuring device
\cite{birrel} is a way
to {\it identify} the event itself, and the particle (or, more
generically, any quantum system) is a derived notion. A quantum system
is then interpreted as a probability amplitude for an event out of a
sequence of events, which takes the place of the ``sequence of
events'' \footnote{In quantum theory, in the absence of trajectories,
  the concept of ``same system'' (i.e.~of a system that persists and
  is re-identifiable in time) is highly problematic \cite{goyal}: the
  GEB formalism reflects this fully, as the system persistence and
  re-identifiability is only enforced as a probability amplitude.},
which is the customary definition of ``physical system'' in relativity.
Schr\"odinger had a similar observation: ``it is better to regard a
particle not as a permanent entity but as an instantaneous event''
\cite{schro}.  Jorge Luis Borges is, obviously, more captivating:
``the world is not a concurrence of objects in space, but a
heterogeneous series of independent acts'' \cite{borges}. Examples of
events (discussed below) are `a fermion with spin $\sigma$ is detected
at a time $t$ and position $\vec x$ in spacetime', or `a boson is
detected with energy $E$ and momentum $\vec p\:$' (more general events
will be discussed elsewhere). While we will be using some of the
formalism developed in \cite{diaz,diaz1}, our interpretation of the
formulas and the conceptual framework of GEB is fundamentally
different.  Relativistic versions of constrained quantum mechanics
\cite{paw,ak,w,qtime,morse,vedral,papersinvedral,rovt,rovelli,montev}
are also explored in
\cite{diaz,diaz1,liebrich,trinity,mehdi,komar1,komar2,gambini,gambinipullin,flaminiarf,piron,piron1,fanchi,stu,stu2,stu3}.
A different claim of `covariant quantum mechanics' is in
\cite{reisenberger}, where the Hilbert space is expressed through
position eigenstates of a time-evolved (Heisenberg picture) position
operator.  While there is some similarity in the notation used, our
approach is completely different: we do not use {\it any} dynamical
assertion (hence, no pictures) in our treatment of time.

The outline follows. We start in Sec.~\ref{sec:hilbert} by introducing
the Hilbert space ${\cal H}_{\bf E}$ for a single-event model,
defining its relativistic observables, giving interpretations of the
system 4D wave-function as unconditioned properties of events and the
representation of Lorentz transformations.  In
Sec.~\ref{sec:multiples} we consider
multiple events in a first quantization approach which is then
embedded it into a second quantization Fock formalism.  In
Sec.~\ref{sec:CORR} we provide a formal connection between GEB and QM
by introducing a correspondence rule that enables one to map quantum
trajectories of the latter into distributions of the former.
Conclusions and future perspectives are given in Sec.~\ref{sec:conc}.
Technical details are presented in appendices.

\section{The Hilbert space of GEB and its canonical observables}\label{sec:hilbert} 
In this section the Hilbert space~${\cal H}_{\bf E}$ which provides the mathematical setting for describing spacetime events is introduced. We start in Sec.~\ref{sec:single} by considering the simplest non trivial scenario, i.e. a universe characterized by a single (spinless) event, which is  the  building block for the general case presented in Sec.~\ref{sec:multiples}. 
 In Sec.~\ref{sec:poincarre} 
we discuss the covariance properties of the theory under Poincar\'e transformations, while in Sec.~\ref{sec:spins} we show how to generalize the analysis to include spinor degrees of freedom.

\subsection{A universe with a single event} \label{sec:single} In the
formulation of GEB we are guided by the fundamental observation that
spacetime is physically meaningful only insofar as it is mapped by
clocks and rods (clicks and ticks) which are events
\cite{geom,science}. From this we can infer that there is no
localization in time of an event without any energy and energy spread
(nothing can happen) and there is no localization in space of an event
without momentum and momentum spread (any event would be delocalized
over the whole space): so, in addition to being characterized by
spacetime coordinates, any given event must also be connected to
energy and momentum degrees of freedom.  Accordingly the structure of
the Hilbert space ${\cal H}_{\bf E}$ associated with a single event
can be identified by declaring that among the linear operators that
acts on such space, there must exist (at least) a 4 component
vectorial observable $\overline{X} := (X^0, X^1,X^2,X^3)$ that
determines the 4-position $\overline{x} := (x^0=t,\vec{x})$ in
spacetime of the event, and an associated 4-momentum operator
$\overline{P} := (P^0, P^1,P^2,P^3)$ that instead defines the
corresponding energy-momentum values $\overline{p} := (p^0=E,\vec{p})$
of the event~\footnote{Notice that at this stage, as it happens in the
  axiomatic definition of QM, we do not provide any physical
  realization of $\overline{X}$ or $\overline{P}$, i.e.~a way to
  detect the spacetime location and momentum of the event. It is
  intuitively clear however that these observables are related with
  the conventional definitions of position and momentum of particles
  in QM: e.g. on one hand if one conditions on the time $t$ of an
  external clock, $\overline{X}$ conditioned on $t$ is just the
  position measurement of a particle happening at some time $t$, the
  conventional position operator of QM that describes a screen that is
  turned on at a certain (externally controlled) time; on the other
  hand, if one conditions on the position $\vec{x}$ of a screen, then
  $\overline{X}$ conditioned on $\vec{x}$ is the time of arrival
  measurement of the particle at an active screen at position
  $\vec{x}$~\cite{mielnik,timeofarrival}. In the non-relativistic
  case, time observables were studied in,
  e.g.,\cite{qtime,pauli,arrival}.}.  (We use the overbar
$\overline{x}$ to denote contravariant 4-vectors, the underbar
$\underline{x}$ for covariant ones, and the arrow $\vec x$ for spatial
3vectors -- see App.~\ref{s:notation}.) The existence of
$\overline{X}$ and $\overline{P}$ is a minimal assumption of the
theory: other observables can in fact be introduced that describe {\it
  extra} (not kinematic) degrees of freedom of the event, something
that for instance will be revealed by the internal degree of freedom
of an event-defined particle (say its spin) (see
Sec.~\ref{sec:spins}).

We now impose the canonical commutation rules
\begin{eqnarray}  
[X^\mu,P^\nu]=-i \eta^{\mu\nu}
  \labell{ccr}\;\mbox{ and }\; [X^\mu,X^\nu]=[P^\mu,P^\nu]=0\;, 
\end{eqnarray}
with $\mu,\nu\in\{0,1,2,3\}$ and $\eta$ the metric $diag(1,-1,-1,-1)$ (note the minus sign in the $00$
commutator).
The rationale of this choice is that, on one hand, it allows us to satisfy 
 Poincar\'e algebra~\cite{diracforms}:
\begin{eqnarray}
[M^{\mu\nu},P^\rho]&=&-i(\eta^{\mu\rho}P^\nu-\eta^{\nu\rho}P^\mu),\qquad
\labell{poinc}\\\nonumber
[M^{\mu\nu},M^{\rho\sigma}]&=&i(\eta^{\nu\rho}M^{\mu\sigma}
-\eta^{\mu\rho}M^{\nu\sigma}\nonumber \\
&& \qquad \qquad -\eta^{\mu\sigma}M^{\rho\nu}+\eta^{\nu\sigma}
M^{\rho\mu}), 
\end{eqnarray}
where $M^{\mu\nu}:=X^\mu P^\nu-X^\nu P^\mu=(X\wedge P)^{\mu\nu}$
\cite{weinberg,hor} is the relativistic angular momentum tensor (the
spatial part
$M^{ij}$ with $i,j\in\{1,2,3\}$, containing the angular momentum tensor, the
generator of rotations, and the temporal part $M^{0j}$ with $j\in\{1,2,3\}$ containing
 the generator of boosts) \footnote{This is the simplest choice to
  satisfy the Poincar\'e algebra, but it is not unique: one can still
  satisfy the commutation relations with appropriate redefinitions of
  $P_\mu$ and $M_{\mu\nu}$.  For example, the instant
  form~\cite{diracforms} arises from the requirement that the position
  and momentum are referred to some instant of time, as in the
  conventional (conditioned) formulation of quantum mechanics
  \cite{orenstein}: in the Schr\"odinger picture the states are
  conditioned to time being $t$ and the operators to ${t}=0$,
  {\it viceversa} in the Heisenberg picture.}. On the other hand, the condition~(\ref{ccr}) automatically imposes   ${\cal H}_{\bf E}$ to be infinite dimensional and leads to the following 
spectral decompositions  
\begin{equation} X^\mu=\int d^4xx^\mu|\overline x\>\<\overline x|\;, \qquad 
P^\mu=\int d^4p\:p^\mu|\overline p\>\<\overline p|\;,\label{defP}\end{equation} 
with $|\overline x\>$ and $|\overline p\>$  forming sets of generalized eigenstates each individually  fulfilling delta-like orthogonality conditions,  i.e. $\<\overline x' |\overline x\>=\delta^{(4)}(
\overline x' -\overline x)$ and $\<\overline p' |\overline p\>=\delta^{(4)}(
\overline p' -\overline p)$, while being related by a 4D
Fourier transform 
\begin{eqnarray} |\overline p\> =\int \tfrac{d^4x}{4\pi^2}e^{-i\overline x\cdot \underline p}|\overline x\>\;. 
  \label{defp} \end{eqnarray} 
(where $\overline{a}=(a^0,\vec{a})$,
$\underline{a}= \eta \overline{a}=(a^0, -\vec{a})$ so that $\overline{a}\cdot \underline{b}=
\overline{a}\eta \overline{b}=a^{\mu} b_{\mu}$, App.~\ref{s:notation})~\footnote{Notice that one can  divide the four 
  degrees of freedom  (1
  temporal and 3 spatial) of  $|\uline{x}\>$  using tensor products
  $|\uline{x}\>=|{t}\>|\vec x\>$, as is done in the nonrelativistic
  Page and Wootters formalism \cite{paw,qtime} (and equivalent ones
  \cite{trinity}) but, this tensor product structure is not absolute to
  preserve Poincar\'e invariance: it is observable-induced
  \cite{paolo,sethpaolo,carroll}, since observers in different reference
  frames will tensorize it differently.}.
The states $|\overline{x}\>$ and $|\overline{p}\>$ are 4D extensions of the usual position and momentum eigenstates $|\vec x\>$, $|\vec p\>$ of QM: just as $|\vec x\>$ and $|\vec p\>$, both 
$|\overline x\>$ and $|\overline p\>$ are not elements of ${\cal H}_{\bf E}$ but 
objects that in functional analysis would correspond to {\it distributions}.
Indeed they belong to the rigged-extended version ${\cal H}_{\bf E}^{+}$ of ${\cal H}_{\bf E}$ that, together with the dense nuclear subspace ${\cal H}_{\bf E}^{-}$ formed by the intersection of the (dense) domains of $\uline{X}$  and $\uline{P}$,
defines the Gelfand  triple 
${\cal H}_{\bf E}^{-} \subset {\cal H}_{\bf E}\subset {\cal H}_{\bf E}^{+}$ of the model. {[The Gelfand  triple \cite{ballentine} is a collection of three objects: the Hilbert space itself  ${\cal H}_{\bf E}$, the set of distributions on the space  ${\cal H}_{\bf E}^{+}$, such as Dirac deltas identifying eigenvectors of continuous-variable eigenvalue observables, and the nuclear elements of the Hilbert space, i.e.~a dense subset ${\cal H}_{\bf E}^{-} $ on which one can apply the distributions.]}
As such, similarly to $|\vec x\>$ and $|\vec p\>$
in QM,  we shall not assign  to them a precise physical interpretation. 
The best we can say is that 
$|\overline x\>$ ($|\overline p\>$) represent an (unphysical) absolute localization of the event in spacetime (resp. energy-momentum space) which can be used  to characterize the statistical distributions of the elements of ${\cal H}_{\bf E}$ via 
decompositions  of the form
 \begin{eqnarray} \label{DEFVECTORS} 
 |\Phi\rangle = \int d^4x\;  \Phi(\overline x)\;  |\overline x\rangle = \int d^4 p \;  
 \tilde{\Phi}(\overline p) \; |\overline p\rangle \;, 
\end{eqnarray}
with the amplitudes 
\begin{eqnarray}
\Phi(\overline x):= \langle \overline x|\Phi\rangle \:, \
\tilde{\Phi}(\overline p):= \langle \overline p |\Phi\rangle = \int \tfrac{d^4x}{4\pi^2}e^{i\overline x\cdot \underline p}
\; \Phi(\overline x)\;, 
\end{eqnarray} being square integrable functions (4D wave-functions)
~\footnote{Formally speaking the decomposition~(\ref{DEFVECTORS}) holds for  the dense subset ${\cal H}_{\bf E}^{-}$ of ${\cal H}_{\bf E}$ which together with
${\cal H}_{\bf E}^{+}$ define the Gelfand triple.}. Quantum probabilities follow then directly from the Born rule. In particular the
probability that the measurement of the observable $X^\mu$ has result
$\overline x$ on a event state associated with the normalized vector $|\Phi\>\in {\cal H}_{\bf E}$ is
\begin{eqnarray}\label{probdistspacetime} 
P(\overline x|\Phi)=|\Phi(\overline{x})|^2= |\<\overline x|\Phi\>|^2\;. \end{eqnarray}  
We interpret this as the 
probability distribution that the event  happens at
spacetime position $\overline x$: this clarifies that 
 in our approach the vectors $|\Phi\rangle$ are 
spacetime states, representing (in the position representation) the
probability amplitudes of the spacetime locations of the event. 
It is worth stressing that in  QM the Born rule $p(\vec x|\psi,{t})=|\<\vec x_S|\psi_S({t})\>|^2=|\<\vec x_H({t})|\psi_H\>|^2$ (written in the Schr\"odinger $S$ or Heisenberg $H$ pictures) gives the {\it conditional} probability that a particle in state $\psi$ is found at position $\vec x$ {\it given} that the time is $t$. Instead, in GEB, the probability \eqref{probdistspacetime} is unconditioned: it is a {\it joint} probability that the event happens at position $\vec x$ {\it AND} that the time is $t$ for $\overline{x}=({t},\vec x)$.
 Analogously, in GEB we can also interpret 
$P(\overline p|\Phi)=|\tilde{\Phi}(\overline{p})|^2=|\<\overline p|\Phi\>|^2$  as the probability that the event
will carry an  energy-momentum~$\overline p$, and  
\begin{eqnarray} p(\Phi_1|\Phi_2)=|\<\Phi_1|\Phi_2\>|^2 \;, \label{probabilities} \end{eqnarray} 
as  the probability to confuse the event $|\Phi_2\rangle$ as the event $|\Phi_1\>
\in {\cal H}_{\bf E}$. 

A direct consequence of  Eq.~\eqref{ccr} are the   Heisenberg-Robertson \cite{robertson}
uncertainty relations 
\begin{eqnarray} \Delta X^\mu\Delta P^\mu\geqslant
  1/2,\label{HRrelations}\end{eqnarray} which for $\mu=0,1,2,3$ have
to be fulfilled by the statistical distributions $P(\overline x|\Phi)$
and $P(\overline p|\Phi)$ for all choices of the normalized vector
$|\Phi\rangle\in{\cal H}_{\bf E}$. These equations effectively
translate in strict mathematical form our initial observation that no
event can be spacetime localized without energy and momentum spread.
While for $\mu=1,2,3$, Eq.~(\ref{HRrelations}) is the usual Heisenberg
uncertainty relation \cite{heis}, for $\mu=0$ it takes a special
role~\cite{matteo,eli}: as a matter of fact it cannot \cite{peresbook}
be derived in QM, where time is a parameter and not an observable.
(Indeed, in the Mandelstamm-Tamm uncertainty relation \cite{man}
$\Delta t$ takes the role of a time interval (the minimum time
interval it takes for a system to evolve to an orthogonal
configuration), not of a statistical uncertainty due to quantum
stochasticity.) Here, instead, time $X^0$ {\it is} a quantum
observable, and we can assign to
$\Delta X^0\Delta P^0\geqslant\hbar/2$ the Heisenberg-Robertson
interpretation. (The acute objections
  \cite{peresbook,aharonovbohm} against this type of interpretation of
  the time-energy uncertainty were formulated in the framework of QM.
  In GEB, these objections do not apply.) This interpretation is a
purely kinematical statement, not a dynamical one: $P^0$ is not some
system's Hamiltonian, but it is the energy devoted to the event. The
connection between the energy and a Hamiltonian (dynamics) is here
only enforced as a constraint on the physical states, see
below.

\subsection{Lorentz tranforms}\label{sec:poincarre} 
Relativistic QM, e.g. QFT,
can be written in a covariant fashion only in very specific
situations. Indeed, as discussed above, the Born rule probability when calculated in the
Schr\"odinger, Heisenberg or interaction picture, is a conditional
probability, conditioned on time. As such, in contrast to what it is claimed
sometimes (e.g.~\cite{halpern}), QM probabilities are
{\it not} covariant: they refer to a specific spacetime foliation, and a
Lorentz transform cannot be simply applied to something defined on a
foliation slice.  In QFT one can recover  covariance if
one writes the observables in the interaction picture in terms of
creation and annihilation operators that have a spacetime dependence
of the form $e^{\pm i\overline{x}\cdot \underline{p}}$. But observables by themselves are
not sufficient to obtain quantum probabilities or expectation values:
they must be applied to states.  QM states explicitly depend on a
foliation (at time $t$ in the Schr\"odinger picture, at time ${t}=0$
in the Heisenberg picture, and at a foliation determined by the
state-evolution operator in the interaction picture). For this reason,
one cannot in general apply a Lorentz transform to a quantum state.
There is an important exception, customarily employed in QFT, where
one uses states that are eigenstates of the Hamiltonian, e.g.~the
vacuum state. These are invariant for the dynamics, and hence can
be easily Lorentz-transformed as they are left invariant.

In contrast, the GEB formalism we introduce here is fully covariant as the canonical observables of the theory are 
trivially transformed using a unitary representation of the Poincar\'e
group. 
To clarify this fact let us consider two inertial observers $O$ and $O'$ sitting on 
different reference frames  $R$ and $R'$  connected via the transformation $\Lambda$ of the Lorentz group so that, 
given  $\overline{x} = (t,\vec{x})$ and $\overline{x}' = (t',\vec{x}')$ the coordinates that they assign to the same event one has
\begin{eqnarray}\label{ddf} 
\overline{x}' = \Lambda \overline{x} \;,
\end{eqnarray} 
(e.g.~if $O'$ has velocity $v$ along the $x$-axis with respect to
$O'$, then $t'= \gamma(t-vx)$, $x' = \gamma (x-vt)$, $y'=y$, and
$z'=z$).  Under these conditions it follows that given a state event
$S$, they will describe it as two different elements
$|\Phi\rangle=\int d^4x \Phi(\overline{x})|\overline{x}\rangle$ and
$|\Phi'\rangle=\int d^4x \Phi'(\overline{x})|\overline{x}\rangle$ of
the space ${\cal H}_{\bf E}$ whose 4D wave-functions amplitudes are
related by the identity
 \begin{eqnarray} 
 \Phi'(\overline{x}) = \Phi(\Lambda^{-1} \overline{x}) \;,\label{phitophi'amp} 
 \end{eqnarray} 
see App.~\ref{S:reference}. 
Accordingly, we can relate the vectors $|\Phi\rangle$ and $|\Phi'\rangle$ as 
\begin{eqnarray} \label{phitophi'}  
|\Phi'\rangle  = U_{\Lambda} | \Phi\rangle \;, 
\end{eqnarray} 
where $U_{\Lambda}$
  is 
 the unitary mapping that represents $\Lambda$ in ${\cal H}_{\bf E}$, i.e. the transformation
 associated with the generators $M^{\mu \nu}$ of Eq.~(\ref{poinc}).
  Equation~(\ref{phitophi'}) ensures that both observers will assign the same scalar products to any two couples of states, i.e. 
\begin{eqnarray} 
\langle \Phi_1'| \Phi_2'\rangle = \langle \Phi_1| \Phi_2\rangle\;,
\end{eqnarray} 
which implies that the probabilities (\ref{probabilities}) are invariant under change of reference frames. 
This is in
line with the fact that the Born rule probabilities in GEB  refer to the
occurrence of events in spacetime (rather than the conditional
probabilities on a foliation at a specific time $t$): they are 
unconditioned and hence invariant. 
As a direct consequence of~(\ref{phitophi'amp}),
 the spacetime probability distributions~(\ref{probdistspacetime}) that $O'$ and $O$ associate with $S$,
 i.e. the functions $P(\overline x|\Phi')=| \Phi'(\overline{x})|^2$ and $P(\overline x|\Phi)=| \Phi(\overline{x})|^2$ respectively, 
 are transformed as any scalar field
\begin{eqnarray}\label{probdistspacetimeO'O} 
P(\overline x|\Phi')=P(\Lambda^{-1} \overline{x}|\Phi)\;. \end{eqnarray}  
Similarly  for the energy-momentum distributions we get  $\tilde{\Phi}'(\overline p)= \tilde{\Phi}(\Lambda \overline p)$ and
hence 
$P(\overline p|\Phi')=P(\Lambda^{-1} \overline{p}|\Phi)$,
(these last follow directly from the identity~(\ref{azionesup}) of  App.~\ref{S:reference}).
A direct consequence of these relations is 
 that any statement $O$ and $O'$  make on the expectation values of these observables on the states of ${\cal H}_{\bf E}$ (or on their
higher momenta like those appearing in~(\ref{HRrelations})) 
are  automatically covariant under Lorentz (and more generally Poincar\'e) transformations. 
This can be made more explicit (App.~\ref{S:reference}) using  the 4D ``Heisenberg picture'' where the mapping~(\ref{phitophi'}) results in the following connection 
between the canonical observables of $O$ and~$O'$
\begin{eqnarray} \label{heisenberg1} 
\overline{X}' &=&  U^\dag_\Lambda \overline{ X}  U_\Lambda  = 
\Lambda \overline{X} \;,\\ \label{heisenberg2} 
\overline{P}' &=&  U^\dag_\Lambda \overline{P}  U_\Lambda  = 
\Lambda \overline{P} \;.
\end{eqnarray}

\subsection{Spinor} \label{sec:spins} As discussed above, extra
(non-kinematic) degrees of freedom can be included into the theory.
This can be done for instance by promoting the generalized
eigenvectors $|\overline x \>$, $|\overline p\>$ to spinor vectors
$|\overline x,\sigma\>$, $|\overline p,\sigma\>$ with $\sigma$ a
spinor index (e.g.~taking values 1,2,3,4) that fulfill generalized
orthogonality conditions
\begin{eqnarray}
\<\overline x',\sigma' |\overline x,\sigma \>&=&\delta_{\sigma,\sigma'} \; \delta^{(4)}(
\overline x' -\overline x)\;,  \nonumber\\
\<\overline p' ,\sigma'|\overline p,\sigma\>&=&\delta_{\sigma,\sigma'}\; \delta^{(4)}(
\overline p' -\overline p)\;.
\end{eqnarray}  Accordingly, the decomposition~(\ref{DEFVECTORS}) of the
event state  is
\begin{equation} \label{DEFVECTORS+spin} 
 |\Phi\rangle = \sum_\sigma \int d^4x\;    \Phi(\overline x,\sigma)\;  |\overline x, \sigma \rangle = \sum_\sigma \int d^4 p \;  
 \tilde{\Phi}(\overline p,\sigma) \; |\overline p,\sigma \rangle, 
\end{equation}
where now the amplitudes $\Phi (\overline x,\sigma):= \langle \overline x,\sigma |\Phi\rangle$ 
(resp.   
$\tilde{\Phi}(\overline p,\sigma):= \langle \overline p,\sigma |\Phi\rangle$) when properly normalized 
define the joint probability $P(\overline x,\sigma|\Phi)=|\Phi(\overline{x},\sigma)|^2$ 
(resp. $P(\overline p,\sigma|\Phi)=|\tilde{\Phi}(\overline{p},\sigma)|^2$) of finding 
$|\Phi\rangle$ in location $\overline{x}$ (with momentum $\overline{p}$) and spin value $\sigma$. 
Under these conditions the identity (\ref{phitophi'}) 
 which relates the description of the observer $O$ and $O'$ still holds by updating (\ref{phitophi'amp}) with 
\begin{eqnarray} 
 \Phi'(\overline{x},\sigma) = \sum_{\sigma'} S^{-1}_{\sigma',\sigma} (\Lambda)\Phi(\Lambda^{-1} \overline{x},\sigma') \;,\label{phitophi'amp+spin} 
 \end{eqnarray} 
 where now $S_{\sigma,\sigma'} (\Lambda)$ is the unitary matrix representation of the Lorentz transform in the spinor space \cite{weinberg}.

 \section{Multiple events} \label{sec:multiples} 
The  Hilbert space ${\cal H}_{\bf E}$ spanned by the vectors~(\ref{DEFVECTORS+spin})  describes a single spacetime event. The extension to the case of multiple events is 
 obtained by considering tensor products  of  such
space, possibly equipped with proper symmetrisation rules 
that account for the statistical properties of the particles that are used to reveal them (see below). 
An important advantage of this construction is that 
the transformations under the Poincar\'{e} group  follow
directly from those established for those of the single-event model~(\ref{phitophi'}).
Indeed  given $|\Phi^{[n]} \rangle,|\Phi^{[n]\prime}\rangle \in {\cal H}^{\otimes n}_{\bf E}$ the vectors used by the observers $O$ and $O'$ to describe the same state of $n$ events, they will be connected via the mapping 
\begin{eqnarray} \label{phitophi'n}  
|\Phi^{[n]\prime}\rangle  = U^{\otimes n}_{\Lambda} |\Phi^{[n]}\rangle \;, 
\end{eqnarray} 
with $U_\Lambda$ being the unitary operator that represents the matrix
$\Lambda$ of Eq.~(\ref{ddf}) for a single event.  Of course with this
choice, each individual event is connected to its own time of
occurrence, whereas interpretations of the covariant formalism in
terms of particles are problematic \cite{diracmultit,diaz} because the
``time of a particle'' is a meaningless concept \cite{peresbook}.  In
the following paragraphs we discuss the different types of multi-event
models that arise from the above formalization and show how this
analysis can be lifted to a higher level of complexity by constructing
an effective quantum field theory version of GEB, through Fock space
(second quantization).

\subsection{Distinguishable vs Indistinguishable events} \label{sec:dist} 
The simplest example of a multi-event model is represented by an universe of 
$n$ {\it distinguishable} spacetime events, i.e. events that define $n$ 
distinguishable particles. In this case  
any normalized vector $|\Phi^{[n]}\rangle$ of  ${\cal H}^{\otimes n}_{\bf E}$ 
qualifies for a proper GEB state of the model,
with the 
decomposition~(\ref{DEFVECTORS+spin}) being replaced by 
\begin{eqnarray} 
 |\Phi^{[n]}\rangle &=& \sum_{\sigma_1,\cdots,\sigma_n} \int d^4x_1 \cdots d^4x_n\;  \Phi^{[n]}(\overline x_1, \sigma_1;\cdots;\overline x_n,\sigma_n
 )\; 
 \nonumber \\&&\qquad \times |\overline x_1,\sigma_1; \cdots ;\overline x_n,\sigma_n\> \label{DEFVECTORSn} \\
  &=& \sum_{\sigma_1,\cdots,\sigma_n} \int d^4p_1 \cdots d^4p_n\;  \tilde{\Phi}^{[n]} (\overline p_1, \sigma_1;\cdots;\overline p_n,\sigma_n
 )\; 
 \nonumber \\&&\qquad \times |\overline p_1,\sigma_1; \cdots ;\overline p_n,\sigma_n\> \;,  \label{DEFVECTORSnP} 
\end{eqnarray}
where $|\overline x_1,\sigma_1; \cdots ;\overline x_n,\sigma_n\>$ stands for $\bigotimes_j|\overline x_j,\sigma_j\>$
with 
 $|\overline x_j,\sigma_j\>_j$  the 4-position and spin eingenstate of the
$j$-th event 
  so that  $\Phi^{[n]}(\overline x_1, \sigma_1;\cdots;\overline x_n,\sigma_n
)$  is the  wave-function which yields the {\it joint} probability
\begin{equation} \label{probn} 
P^{[n]}(\overline x_1, \sigma_1;\cdots;\overline x_n,\sigma_n
) = |\Phi^{[n]}(\overline x_1, \sigma_1;\cdots;\overline x_n,\sigma_n
)|^2\;, 
\end{equation} 
 of  revealing the $j$th event in spacetime position $\overline x_j$ with spin $\sigma_j$, for all $j$. ---  similar definitions apply also to 
 $|\overline p_1,\sigma_1; \cdots ;\overline p_n,\sigma_n\>$ and $\tilde{\Phi}^{[n]}(\overline p_1, \sigma_1;\cdots;\overline p_n,\sigma_n
 )$ of Eq.~(\ref{DEFVECTORSnP}).

 Consider next the scenario of $n$ {\it indistinguishable} spacetime events, i.e. 
 events that are used to define
 $n$ identical particles. We
  describe the states of such models 
 via vectors~(\ref{DEFVECTORSn}) which induce the Bosonic or Fermionic character of the derived particles through the property of being either completely symmetric
 or completely anti-symmetric  under exchange of the ket indexes.
Specifically a Bosonic $n$-event   GEB model is described by the completely symmetric linear subset ${\cal H}^{(n,{\bf S})}_{\bf E}\subset {\cal H}^{\otimes n}_{\bf E}$ spanned by the vectors 
~(\ref{DEFVECTORSn})
with  amplitudes probabilities $\Phi^{[n]}(\overline x_1, \sigma_1;\cdots;\overline x_n,\sigma_n)$ that are invariant under an arbitrary permutation ${\mathbf{p}}$ of the $n$ systems labels, i.e.
\begin{eqnarray}  \label{symmBOS} 
&&\Phi^{[n]}(\overline x_{{\mathbf{p}}(1)},\sigma_{{\mathbf{p}}(1)}; \cdots;\overline x_{{\mathbf{p}}(n)},\sigma_{{\mathbf{p}}(n)}
 )\\
 && \qquad\qquad \qquad \qquad = \nonumber 
\Phi^{[n]}(\overline x_1, \sigma_1;\cdots;\overline x_n,\sigma_n)\;, \quad \forall {\mathbf{p}}\;, \end{eqnarray} 
(a condition that automatically carries over to the 4D momentum wave-function).  
  A Fermionic   $n$-event   GEB model, instead, will be described by the
 completely anti-symmetric linear subset 
  ${\cal H}^{(n,{\bf A})}_{\bf E}\subset {\cal H}^{\otimes n}_{\bf E}$ formed by vectors with 
 amplitudes that fulfill the identity  
\begin{eqnarray}&& \Phi^{[n]} (\overline x_{{\mathbf{p}}(1)},\sigma_{{\mathbf{p}}(1)}; \cdots;\overline x_{{\mathbf{p}}(n)},\sigma_{{\mathbf{p}}(n)}  \label{symmFER} 
 )\\
 && \qquad\qquad \qquad =\mbox{sign}[{\mathbf{p}}] \; \nonumber 
 \Phi^{[n]} (\overline x_1, \sigma_1;\cdots;\overline x_n,\sigma_n)\;,\quad \forall {\mathbf{p}}\;,  \end{eqnarray}  
with $\mbox{sign}[{\mathbf{p}}]$ being the sign of the permutation ${\mathbf{p}}$.

\subsection{Fock space representation} \label{FOCK} 
The above construction can only deal with a fixed, predetermined number $n$ of events. To analyze situations where $n$ is, itself, a quantum degree of freedom, we need to escalate to a Fock space.
The starting point of this construction is to introduce 
 a ``4D-vacuum'' state vector $|0\>_4$ which represents the state of a universe where
there are no events {\it at any spacetime location}, and by defining raising and lowering
operators \cite{diaz,liebrich} $a^\dag_{\overlinelz x,\sigma}$, $a_{\overlinelz x,\sigma}$ that   act as creators/annihilators
of spacetime events in the  theory. 
Properly symmetrized versions of the generalized 4-position  eigenstates  
will now be expressed as reiterated applications of the $a^\dag_{\overlinelz x,\sigma}$'s on $|0\>_4$, i.e.
\begin{equation} \label{defnewn}  
|\overline x_1,\sigma_1;\cdots ;\overline x_n,\sigma_n\> \quad 
\mapsto  \quad \frac{1}{\sqrt{n!}} \; 
a^\dag_{\overlinelz
  x_1, \sigma_1 }\cdots a^\dag_{\overlinelz x_n, \sigma_n}|0\>_4\;,
\end{equation} 
while 
the generalized 4-momentum eigenstates $|\overline p_1,\sigma_1;\cdots ;\overline p_n,\sigma_n\> $ as \begin{equation}  \label{repp} 
|\overline p_1,\sigma_1;\cdots ;\overline p_n,\sigma_n\> \quad 
\mapsto  \quad \frac{1}{\sqrt{n!}} \;  a^\dag_{\overlinelz
  p_1, \sigma_1 }\cdots a^\dag_{\overlinelz p_n, \sigma_n}|0\>_4\;,
\end{equation} 
with $a^\dag_{\overlinelz p,\sigma}$  and $a_{\overlinelz p,\sigma}$  connected with 
$a^\dag_{\overlinelz x,\sigma}$, i.e. the operators 
\begin{equation}  \label{dend} 
a^\dag_{\overlinelz p,\sigma} := \int
\tfrac{d^4x}{4\pi^2}e^{-i\overlinelz p \cdot \underline{x}}  a^\dag_{\overlinelz x,\sigma}\;,
\qquad 
a_{\overlinelz p,\sigma} := \int
\tfrac{d^4x}{4\pi^2}e^{i\overlinelz p \cdot \underline{x}}  a_{\overlinelz x,\sigma}\;.
\end{equation} 
The consistency of the representation is enforced by assigning proper commutation relations to the
raising and lowering operators (App.~\ref{app:multi}). Specifically, 
the
 Bosonic/Fermionic character follows by requiring 
\begin{align} 
  \mbox{Bose: } [a_{\overlinelz x, \sigma},a^\dag_{\overlinelz
  x', \sigma'}]=\delta_{\sigma,\sigma'} \delta^{(4)}(\overlinelz
  x-\overlinelz x'), \; 
   [a_{\overlinelz x, \sigma},a_{\overlinelz
  x', \sigma'}]=0, \ \ \:
\nonumber
\\
 \mbox{Fermi:}  \{a_{\overlinelz x, \sigma},a^\dag_{\overlinelz
  x', \sigma'}\}=\delta_{\sigma,\sigma'} \delta^{(4)}(\overlinelz
  x-\overlinelz x'),\; \{a_{\overlinelz x, \sigma},a_{\overlinelz
  x', \sigma'}\}=0, \label{ccr1}
\end{align}
which automatically translate into analogous relations for the $a^\dag_{\overlinelz p,\sigma}$, $a_{\overlinelz p,\sigma}$, i.e.
\begin{align} 
  \mbox{Bose: } [a_{\overlinelz p, \sigma},a^\dag_{\overlinelz
  p', \sigma'}]=\delta_{\sigma,\sigma'} \delta^{(4)}(\overlinelz
  p-\overlinelz p'), \; 
   [a_{\overlinelz p, \sigma},a_{\overlinelz
  p', \sigma'}]=0, \ \ \:
\nonumber\\
\label{ccr12}
 \mbox{Fermi:}  \{a_{\overlinelz p, \sigma},a^\dag_{\overlinelz
  p', \sigma'}\}=\delta_{\sigma,\sigma'} \delta^{(4)}(\overlinelz
  p-\overlinelz p'),\; \{a_{\overlinelz p, \sigma},a_{\overlinelz
  p', \sigma'}\}=0. 
\end{align}
Accordingly in the Fock state representation~Eqs.~(\ref{DEFVECTORSn}) and (\ref{DEFVECTORSnP}) 
can be expressed as
\begin{eqnarray}
\!\!\!|\Phi^{[n]}\>\!\!\!&=&\!\!\! \frac{1}{\sqrt{n!}}  \sum_{\sigma_1,\cdots,\sigma_n} \int d^4x_1\cdots
  d^4x_n\:\Phi^{[n]} (\overline x_1, \sigma_1;\cdots;\overline x_n,\sigma_n
 )\nonumber \\
  &&\qquad \qquad \times \;  a^\dag_{\overlinelz x_1,\sigma_1}\cdots
  a^\dag_{\overlinelz x_n,\sigma_n}|0\>_4 
\labell{psiposr} \\
\!\!\!&=&\!\!\!\frac{1}{\sqrt{n!}}  \sum_{\sigma_1,\cdots,\sigma_n} \int d^4p_1\cdots
  d^4p_n\:\tilde{\Phi}^{[n]}(\overline p_1, \sigma_1;\cdots;\overline p_n,\sigma_n
 )\nonumber \\
  &&\qquad \qquad \times \;  a^\dag_{\overlinelz p_1,\sigma_1}\cdots
  a^\dag_{\overlinelz p_n,\sigma_n}|0\>_4 \labell{psipospene} \;, 
\end{eqnarray}
with $\Phi^{[n]} (\overline x_1, \sigma_1;\cdots;\overline x_n,\sigma_n
 )$ and $\tilde{\Phi}^{[n]}(\overline p_1, \sigma_1;\cdots;\overline p_n,\sigma_n
 )$
  retaining the same probabilistic meaning given in Eq.~(\ref{probn}).
 
  It is important to stress that the 4D-vacuum $|0\>_4$ is a distinct
  state from the 3D-vacuum $|0\>_3$ used in QFT. Indeed $|0\>_3$
  represents a configuration in which there are no particles {\it at a
    specific time} (time $t$ in the Schr\"odinger picture or ${t}=0$
  in the Heisenberg one), whereas $|0\>_4$ has no events at any time.
 The QFT vacuum $|0\>_3$ is the ground state of a
    field. As such, it is the spatial part (in some foliation) of the
    GEB state relative to a zero 4-momentum event:
    $|0\>_3=$foliate$(a^\dag_{\overline{p}=0}|0\>_4)$. It is {\it not}
    the spatial part of $|0\>_4$. The state $|0\>_4$ represents no
    events, whereas $a^\dag_{\overline{p}=0}|0\>_4$ represents a zero
    4-momentum event which corresponds to a uniform distribution of
    zero-energy events in all spacetime
    $a^\dag_{\overline{p}=0}|0\>_4=\int
    d^4x\:a^\dag_{\overline{x}}|0\>_4$: in other words, there is a
    difference between saying ``nothing happens everywhere and
    everywhen'' (i.e.~the QFT vacuum $|0\>_3$ at all times), and
    saying ``nothing happens anywhere and at any time'', the
    Aristotelian void $|0\>_4$.

    Also, the GEB raising and lowering operators are completely
    different from the QFT ones, and it is not just a matter of adding
    the temporal degree of freedom: $a_{\overline{p}}\neq a_{p^0}\otimes a_{\vec p}$. The QFT ones
    are obtained from the canonical quantization of the harmonic
    oscillator, namely starting from the {\it dynamics}. Here,
    instead, we are introducing the raising and lowering operator from
    the {\it kinematics}, namely the ones that applied to the 4-vacuum
    create the position (or momentum) eigenstates.  The QFT operators
    lose their meaning when one changes the dynamics (e.g. by adding
    interactions), whereas ours do not.

Thanks to Fock space, we can now have states with superpositions of
different numbers of events, i.e.
\begin{eqnarray}\label{superPP} 
|\Phi\>&=&\sum_{n\geq 0} \alpha_n |\Phi^{[n]}\rangle \;, 
\end{eqnarray}
with $|\Phi^{[0]}\rangle := |0\rangle_4$ and  $\alpha_n$ probability amplitudes, so that
$|\alpha_n|^2  |\Phi^{[n]}(\overline x_1, \sigma_1;\cdots;\overline x_n,\sigma_n
)|^2$
   (resp. $|\alpha_n|^2 |\tilde{\Phi}^{[n]}(\overline p_1, \sigma_1;\cdots;\overline p_n,\sigma_n
 )|^2$) is the joint  probability density of having
$n$ detection events {\it and} that they happen at spacetime locations
$\overlinelz x_1,\cdots,\overlinelz x_n$ (resp. momenta $\overlinelz p_1,\cdots,\overlinelz p_n$) and with spins $\sigma_1,\cdots, \sigma_n$ ($|\alpha_0|^2$ being the probability of no event). 

The Lorentz transformations are a straightforward extension of 
(\ref{phitophi'n}). Indeed, indicating 
with ${\cal U}_{\Lambda}$ the unitary mapping that represents $\Lambda$ in Fock space, given $|\Phi\rangle$
and $|\Phi'\rangle$ the states two observers $O$ and $O'$ assign to same state event, we can
write 
\begin{eqnarray} \label{phitophi'nfock}  
|\Phi'\rangle  = {\cal U}_{\Lambda} | \Phi\rangle \;, 
\end{eqnarray}
by requiring that the vacuum state is left invariant by
${\cal U}_{\Lambda}$, ${\cal U}_{\Lambda} |0\rangle_4= |0\rangle_4$,
and by imposing
 \begin{eqnarray} \label{nice1} 
 {\cal U}_{\Lambda} a^\dag_{\overlinelz x,\sigma} {\cal U}^\dag_{\Lambda}=
 \sum_{\sigma'} S_{\sigma,\sigma'}^{-1}(\Lambda)\;  a^\dag_{\Lambda \overlinelz x,\sigma'}\;,
 \end{eqnarray} 
 or, equivalently,
 \begin{eqnarray}\label{nice2} 
 {\cal U}_{\Lambda} a^\dag_{\overlinelz p,\sigma} {\cal U}^\dag_{\Lambda}=
 \sum_{\sigma'} S_{\sigma,\sigma'}^{-1}(\Lambda) \; a^\dag_{\Lambda \overlinelz p,\sigma'}\;. 
 \end{eqnarray}

\section{QM/GEB correspondence}\label{sec:CORR}

QM is a physical theory of systems while GEB is a physical theory of events: 
in this section we shall see that  these two different approaches can be connected by 
associating 
the dynamical quantum trajectories of QM to elements of the  distributions set  ${\cal H}^+_{\bf E}$ of  GEB.
 For the sake of simplicity we start in Sec.~\ref{sec:primasec} by considering  the  special case of a single event universe
  showing that it can be put in correspondence with the QM description of a point-like single particle system.
  The generalization to multi-event scenario will instead be addressed in Sec.~\ref{sec:MULTI} and in 
  Sec.~\ref{sec:NONINT} where we shall make use of the 
 Fock space representation introduced in Sec.~\ref{FOCK}.

\subsection{Single-particle/single-event correspondence} \label{sec:primasec} 
In QM the temporal evolution of a single particle described by 
an observer $O$ sitting in his reference frame $R$, 
is obtained  by assigning a  3D+1 spinor wave-function  $\Psi_{\rm QM}(\vec x,\sigma|t)$ which extends
both in time 
and in
space.
The unitary character of the  dynamics ensures that 3D norm of this function is a constant of motion. Accordingly
setting 
  \begin{eqnarray}\label{normaQM} 
 \sum_\sigma \int d^3 x |\Psi_{\rm QM}(\vec x,\sigma|t)|^2=1\;,\qquad \forall t\in \mathbb{R} \end{eqnarray} 
the function  
$\Psi_{\rm QM}(\vec x,\sigma|t)$ can be interpreted as
the conditional probability amplitude that  the observer $O$ will find the particle   at
position $\vec x$ with spin~$\sigma$, given that time is $t$.
 A natural  correspondence between the single particle states of QM and the single event states of the GEB formalism can hence be established by 
 interpreting $\Psi_{\rm QM}(\vec x,\sigma|t)$ as a 4D spinor wave-function 
 \begin{eqnarray} \Psi_{\rm QM}(\overline x,\sigma):=\Psi_{\rm QM}(\vec x,\sigma|t)\;, \label{ideddit} \end{eqnarray} 
 and then using the 
following mapping
 \begin{eqnarray}\labell{las}
 &&\Psi_{\rm QM}(\overline{x},\sigma) \in {\rm QM} 
\\ \nonumber 
 &&\qquad \qquad \mapsto\;   |\Psi_{\rm QM}\>:\stackrel{!}{=}\sum_\sigma
  \int d^4x\; \Psi_{\rm QM}(\overline
  x,\sigma)\;|\overline x,\sigma\>\;.
\;
\end{eqnarray}
The exclamation mark is a reminder
that, with the normalization \eqref{normaQM}, the vector
$|\Psi_{\rm{QM}}\>$ has an infinite norm, in contrast to $|\Phi\>$ of
\eqref{DEFVECTORS}. Indeed the square integrability of the GEB
wave-functions $\Phi(\overline{x},\sigma)$ is incompatible with
(\ref{normaQM}) obeyed by the QM wave-function
$\Psi_{\rm QM}(\overline{x},\sigma)$: in general an element $|\Phi\>$
of ${\cal H}_{\bf E}$ will exhibit modulations with respect to $t$
that in QM would be interpreted as unphysical {\it losses} and {\it
  gains} of probability during the temporal evolution of the particle
but which are perfectly allowed at the kinematic level in the GEB
formalism (and they can then be removed at the dynamical level, see below).

Because of their infinite norm, the vectors $|\Psi_{\rm{QM}}\>$
introduced above are not elements of ${\cal H}_{\bf E}$ and cannot be
interpreted as proper event states of GEB.  The mapping (\ref{las})
 associates the QM wave-functions
$\Psi_{\rm QM}(\vec x,\sigma|t)$ of $O$ to distributions of GEB. This
fact is explicitly shown in App.~\ref{s:QMGEBcorrespondence}: here
 we point out that Eq.~(\ref{las}) identifies only a proper
subset ${\cal H}_{\bf QM}$ of ${\cal H}^+_{\bf E}$. Examples of
elements of ${\cal H}^+_{\bf E}$ which are not in ${\cal H}_{\bf QM}$
are provided for instance by the generalized position and momentum
eigenvectors $|\overline{x}\rangle$ and $|\overline{p}\rangle$ which
clearly cannot be expressed as in ~(\ref{las}) with 3D normalized QM
solutions $\Psi_{\rm QM,\sigma}(\vec{x}|t)$.  We also notice
${\cal H}_{\bf QM}$ can be identified via geometric constraints
analogous to those adopted
in~\cite{wdw,paw,hor,hor1,diaz,diaz1,rovellibook,trinity,halliwel,gambini}.
Specifically one has
\begin{eqnarray} \label{CONSTRAINT1} 
|\Psi_{{\rm QM}}\> \in {\cal H}_{\bf QM} \quad \Longleftrightarrow \quad 
 \left\{ \begin{array}{l} K |\Psi_{{\rm QM}}\>=0\;, \\ \\
 |\Psi_{{\rm QM}}\> \neq 0 \;, 
 \end{array} \right.
\end{eqnarray}
where $K$ is a constraint operator that encodes the QM dynamics (as
discussed in App.~\ref{secinitial} we can also add extra constrains
that force $|\Psi_{\rm QM}\rangle$ to represent 3D+1 spinor
wave-functions that fulfill assigned initial conditions for a given
observer $O$).  We stress that in contrast to previous literature
\cite{diaz,diaz1,piron1,hor} where the solutions of constraint
equations are interpreted as history states for {\it systems}, in GEB
they are used to identify distributions which define 
{\it event} states. Another difference with previous approaches is
that for the purpose to generalizing the analysis to the multi-event
scenarios, in our construction we find it convenient to work with
constraint operators which are explicitly self-adjoint and
semidefinite-positive, i.e.  $K \geq 0$.  Of course this choice can be
enforced without loss of generality since given $J$ a generic operator
fulfilling~(\ref{CONSTRAINT1}) we can always identify a positive
semi-definite one that does exactly the same e.g. by taking
$K=J^\dag J$ exploiting the fact that
\begin{eqnarray} 
J |\Psi_{{\rm QM}}\>=0 \qquad  \Longleftrightarrow \qquad J^\dag J |\Psi_{{\rm QM}}\>=0\;. \end{eqnarray}

For non-relativistic models, the QM dynamics takes always the form of
a Schr\"odinger equation which can be cast in the
form~(\ref{CONSTRAINT1}) along the lines detailed e.g. in in
Ref.~\cite{qtime}. Unfortunately, there does not appear to be a
similarly general method to describe the relativistic dynamics
\footnote{A guideline, suggested by Wigner and Bargmann, is to
  consider as physical fields the ones that correspond to irreducible
  representations of the Poincar\'e group \cite{weinberg}.}, but one
should use covariant constraints to avoid ruining the theory's
covariance.  Indicating with $\square:= \partial^2_t - \nabla^2$ the
D'Alambert operator, in the case of spinless particle of mass $m$ this
can be done for instance by invoking the Klein-Gordon (KG) equation
\begin{eqnarray}\label{KGEQ} (\square +m^2)\Psi_{{\rm
      QM}}(\vec{x}|t)\Big|_{+}=0 \;, \end{eqnarray} filtering out its
positive energy solutions (see App.~\ref{KGdynamics}). For the case of
a massive spin 1/2 particle instead one can use the Dirac equation
\begin{eqnarray} \label{DIRAC} 
\sum_{\sigma=1}^4(i\overline{\gamma}_{\sigma',\sigma}\cdot \underline{\partial} - m\delta_{\sigma',\sigma})\Psi_{{\rm QM}}(\vec{x},\sigma|t)=0\;,
\end{eqnarray}  
with $\overline{\partial} := (\partial/\partial {t},-\vec\nabla)$ and
$\overline{\gamma}:=(\gamma^0,\gamma^1,\gamma^2,\gamma^3)$ the Dirac
matrices (see Eq.~(\ref{gammadef})). Adopting the position
representation $P^\mu\to i\partial^\mu$, both Eqs.~(\ref{KGEQ}) and
(\ref{DIRAC}) can be turned into a constraint of the form
(\ref{CONSTRAINT1}) for the associated distributions
$|\Psi_{\rm QM}\rangle$ {\footnote{ An interaction with an
  external electromagnetic field can be described through the
  minimal coupling substitution of $P^\mu$ with $P^\mu+eA^\mu$, with
  $e$ the particle charge and $A^\mu$ the em 4potential.)}}.
Specifically, in the case of the positive-energy KG
equation~(\ref{KGEQ}) one can identify the constraint operator $K$
of~(\ref{CONSTRAINT1}) with the self-adjoint operator
 \begin{equation}
J_{\rm KG^{+}} := \int d^4p \; ( \Theta( p^0) \; \overline{p} \cdot \underline{p} - m^2)
  |\overline{p}\rangle \langle \overline{p}|\;, 
\labell{constr+} 
\end{equation}
where  $\Theta(x)$  is  the Heaviside step function 
 (see App.~\ref{KGdynamics}) or with its positive definite counterpart 
  \begin{equation}
K_{\rm KG^{+}}:= J^2_{\rm KG^{+}} = \int d^4p \; \Big( \Theta( p^0) \; \overline{p} \cdot \underline{p} - m^2 \Big)^2
  |\overline{p}\rangle \langle \overline{p}|\;.
\labell{constr++} 
\end{equation}

\begin{figure}
	\includegraphics[width=\columnwidth]{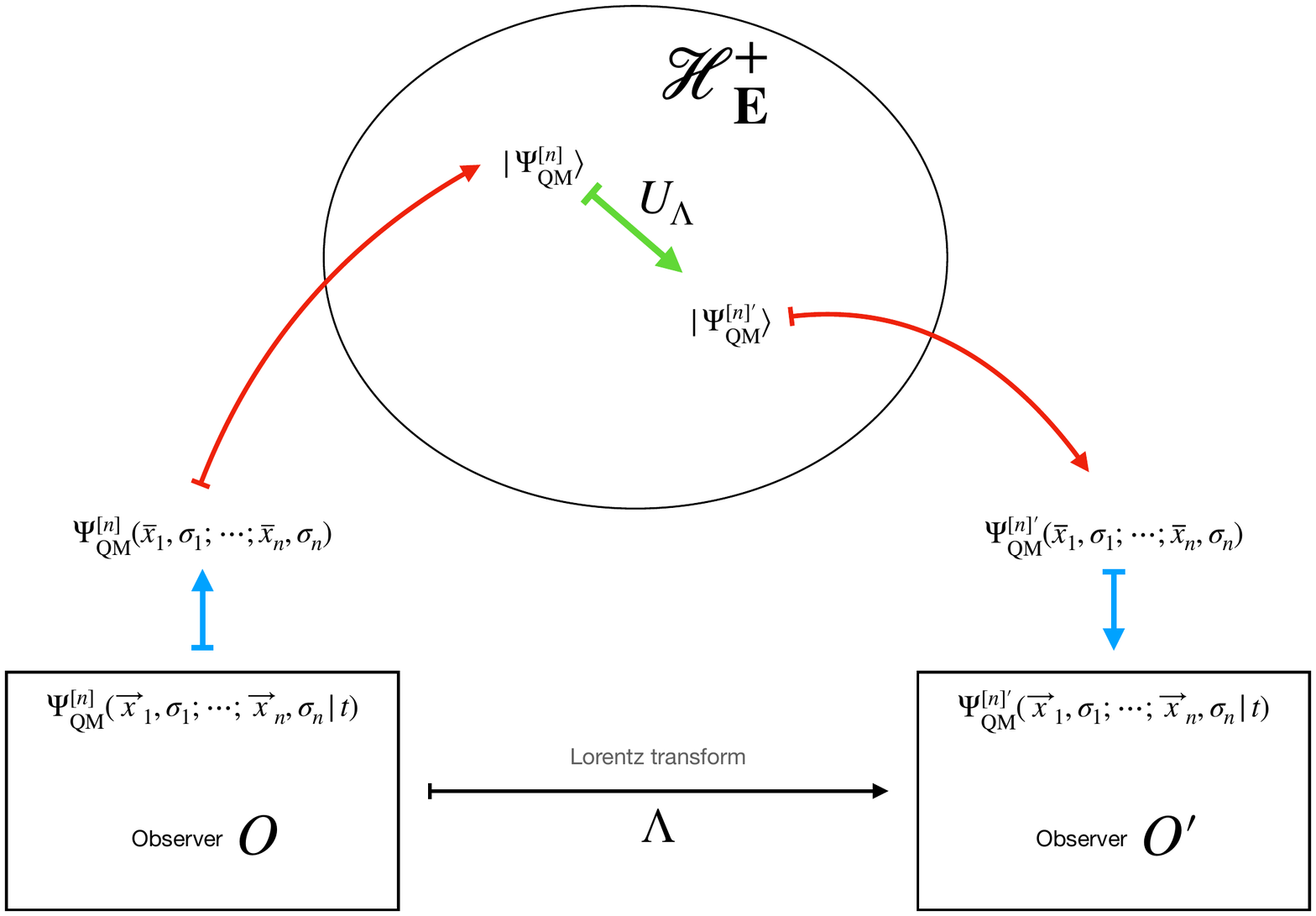}
	\caption{Schematic representation of the connection that links
          the 3D+1 spinor wave-functions
          $\Psi^{[n]}_{\rm QM}(\vec{x}_1,\sigma_1;\cdots;
          \vec{x}_n,\sigma_n |t)$ and
          $\Psi^{[n]\prime}_{\rm QM}(\vec{x}_1,\sigma_1;\cdots;
          \vec{x}_n,\sigma_n|t)$ of QM, that two observers $O$ and $O'$
          assign to an $n$-particle state in their own
          reference frames: thanks to the QM/GEB
          correspondence~(\ref{las}) such connection can be expressed
          via the unitary mapping~(\ref{did}) (green arrow) that links
          the associated GEB distributions~(\ref{lasgen}).  The blue arrow
          elements represent the association of 3D+1 spinor
          wave-functions of QM with their 4D spinor
          GEB counterparts, given in~(\ref{ideddit}) for $n=1$ and
          (\ref{chiave}) for $n>1$; the red arrow elements instead
          represent the connection between 4D spinor wave-functions
          and the distributions of ${\cal H}_{\rm E}$. }
	\label{figurecorrespondence}
\end{figure}

Similarly, for  the Dirac equation: we can directly translate (\ref{DIRAC}) into (\ref{CONSTRAINT1}) 
by identifying $K$ with the operator
\begin{eqnarray}  \labell{constrDIR}
J_{\rm D}&:=& \overline{\gamma} 
\cdot \underline{P} - m \\ \nonumber 
&=&
\sum_{\sigma,\sigma'} \int d^4p \; (\overline{\gamma}_{\sigma,\sigma'} \cdot \underline{p} - m\; \delta_{\sigma,\sigma'}) |\overline{p},\sigma \rangle \langle \overline{p},\sigma'| \;, 
\end{eqnarray}
(which is  not self-adjoint), 
or with its associated positive semi-definite counterpart 
\begin{equation}  \labell{constrDIR++}
K_{\rm D}:= J_{\rm D}^\dag J_{\rm D} =
\sum_{\sigma=1}^4 \int d^4p \; \lambda^2_\sigma (\overline{p}) \;  |\phi_\sigma (\overline{p})\rangle \langle \phi_\sigma (\overline{p})| \;, 
\end{equation}
with $|\lambda_\sigma (\overline{p})|$  being the singular eigenvalues of $J_{\rm D}$ 
and the generalized  vectors $|\phi_\sigma (\overline{p})\rangle$ forming an orthonormal set
\begin{eqnarray}\label{neworthonw} 
\<\phi_{\sigma'} (\overline{p}^\prime)  |\phi_\sigma (\overline{p}) \rangle&=&\delta_{\sigma,\sigma'}\; \delta^{(4)}(
\overline p\;^\prime -\overline p)\;,
\end{eqnarray} 
(see App.~\ref{appconstraintDIRAC}).
One of the  advantages of adopting the above definitions
 for the constraint operator  $K$ is  that all of them 
are  Lorentz invariant quantities (this is clearly evident for  (\ref{constrDIR}), while for
 (\ref{constr+}) an explicitly  proof is given in  App.~\ref{KGdynamics}). Accordingly,  the elements of ${\cal H}_{\bf QM}$ identified by one observer $O$ via Eq.~(\ref{CONSTRAINT1})  will be related with
those assigned by the observer $O'$ via the same unitary transformation~(\ref{phitophi'}) that links their
state event descriptions, i.e.
\begin{eqnarray}\label{did} 
|\Psi'_{\rm QM}\rangle = U_{\Lambda} |\Psi_{\rm QM}\rangle\;,  
\end{eqnarray} 
or equivalently 
\begin{eqnarray} \label{didnot} 
\Psi'_{\rm QM} (\overline{x},\sigma) &=& \sum_{\sigma} S_{\sigma',\sigma}^{-1}(\Lambda) \Psi_{\rm QM} (\Lambda^{-1} \overline{x},\sigma')\;, 
\end{eqnarray} 
(the spinless case  being obtained by simply removing $S$ and neglecting the $\sigma$ terms),  
which via (\ref{ideddit}) 
properly describes how to relate the QM 3D+1 spinor wave-functions 
$\Psi'_{\rm QM} (\vec{x},\sigma|t)$ and $\Psi_{\rm QM} (\vec{x},\sigma|t)$ 
the observers assign to the {\it same}
single-particle trajectory (see Fig.~\ref{figurecorrespondence}). 

\subsection{Multi-event QM/GEB correspondence} \label{sec:MULTI} 

To  generalize the correspondence~(\ref{las}) to the multi-event case we need to address the problem that
in QM the wave-function $\Psi^{[n]}_{\rm QM}(\vec{x}_1,\sigma_1; \cdots; \vec{x}_n, \sigma_n|t)$ of a $n$ particle system is  associated with $n$ independent 3D spatial coordinates (plus possibly $n$ spinor components) but with a single time-coordinate. It is hence not at all clear how to map such terms into  elements (or distributions) of
 ${\cal H}_{\rm E}^{\otimes n}$ which instead possess $n$ independent time coordinate values. 
 In the case where the $n$ particles are not interacting, we can use the fact that $\Psi^{[n]}_{\rm QM}(\vec{x}_1,\sigma_1; \cdots; \vec{x}_n, \sigma_n|t)$ can always be expressed as linear combinations of  products of time-dependent single-particles, i.e.
\begin{eqnarray}\label{chiave} 
&&\Psi^{[n]}_{\rm QM}(\vec{x}_1,\sigma_1; \cdots;  \vec{x}_n, \sigma_n|t) \\ \nonumber 
&&\qquad \qquad = \sum_{\vec{\ell}} \alpha_{\vec{\ell}} \; 
\Psi^{(\ell_1)}_{\rm QM}(\vec{x}_1,\sigma_1| t) \cdots \Psi^{(\ell_n)}_{\rm QM}(\vec{x}_n, \sigma_n|t)\;, 
\end{eqnarray} 
where  given $\vec{\ell} = (\ell_1,\cdots,\ell_n)$, $\alpha_{\vec{\ell}}$ are time-independent probability amplitudes,
and where for $j=1, \cdots, n$, $\Psi^{(\ell_j)}_{\rm QM}(\vec{x}_j, \sigma_j|t)$ is the 3D+1 wave-function describing the evolution of the $j$-th particle of the system.   
Equation~(\ref{chiave}) is the key to generalize (\ref{las}) as it allows us to
formally associate $\Psi^{[n]}_{\rm QM}(\vec{x}_1,\sigma_1; \cdots; \vec{x}_n, \sigma_n|t)$ to
a 4D spinor wave-function with $n$ distinct time coordinates via the construction 
\begin{eqnarray} \label{imoimpo} 
&&\Psi^{[n]}_{\rm QM}(\overline{x}_1,\sigma_1; \cdots;  \overline{x}_n, \sigma_n)\\ \nonumber 
&& \qquad \quad  := 
 \sum_{\vec{\ell}} \alpha_{\vec{\ell}} \; 
\Psi^{(\ell_1)}_{\rm QM}(\vec{x}_1,\sigma_1| t_1) \cdots \Psi^{(\ell_n)}_{\rm QM}(\vec{x}_n, \sigma_n|t_n)\;,
\end{eqnarray} 
and then using such term to identify the distribution $|\Psi_{\rm QM}\>$ via the identity 
 \begin{eqnarray}\labell{lasgen}
 && |\Psi^{[n]}_{\rm QM}\>:\stackrel{!}{=}\sum_{\sigma_1,\cdots, \sigma_2} 
  \int d^4x_1\cdots  \int d^4x_n \\ \nonumber 
  && \qquad \quad \times \; \Psi^{[n]}_{\rm QM}(\overline{x}_1,\sigma_1; \cdots;  \overline{x}_n, \sigma_n)\; 
  |\overline{x}_1,\sigma_1; \cdots;  \overline{x}_n, \sigma_n\rangle  
\;
\end{eqnarray}
(where, again, ``$!$'' is a reminder of the non-normalization).
While the choice of the single-particles QM spinor wave-functions
$\Psi^{(\ell_j)}_{\rm QM}(\vec{x}_j,\sigma_j| t)$ and of the coefficients $\alpha_{\vec{\ell}}$ entering in~(\ref{chiave}) are in general not unique,
the vector 
(\ref{lasgen}) does not depend on such freedom ensuring that  the connection between 
$\Psi^{[n]}_{\rm QM}(\vec{x}_1,\sigma_1; \cdots; \vec{x}_n, \sigma_n|t)$ and $|\Psi^{[n]}_{\rm QM}\>$ is one-to-one 
(see App.~\ref{app:uniq}). {\it Viceversa}, given $|\Psi^{[n]}_{\rm QM}\>$ one can recover the QM spinor 3D wave-function 
$\Psi^{[n]}_{\rm QM}(\vec{x}_1,\sigma_1; \cdots; \vec{x}_n, \sigma_n|t)$ via the identity 
\begin{eqnarray} \label{contratime} 
&&\Psi^{[n]}_{\rm QM} (\vec{x}_1,\sigma_1;\cdots; 
\vec{x}_n,\sigma_n|t)\\ \nonumber 
&& \qquad \qquad =\langle \overline{x}_1, \sigma_1 ;\cdots;
\overline{x}_n, \sigma_n|\Psi^{[n]}_{\rm QM}\rangle\Big|_{t_1=\cdots=t_n=t} \;. 
 \end{eqnarray}  
As in the single-event case we can identify the distributions (\ref{lasgen}) by means of a geometric constraint~(\ref{CONSTRAINT1}) induced by an $n$-body operator $K^{[n]}$. 
To identify such a term we start form 
 individual single particle constraint terms $K_j$ that are explicitly positive semidefinite 
(i.e. $K_j\geq 0$), and 
 take $K^{[n]}$ as their  sum 
 \begin{eqnarray}
 K^{[n]} = \sum_{j=1}^n K_j \;,  \label{PPPP} \qquad \qquad (K_j\geq 0)\;. 
 \end{eqnarray} 
The  positivity requirement on the individual $K_j$  is an important ingredient  as it ensures that 
 the kernel of $K^{[n]}$ coincides with the intersection of all the kernels of the single-particle constraints, i.e.  
\begin{equation}
\label{papq}
K^{[n]}|\Psi^{[n]}_{\rm QM}\rangle =0 \Leftrightarrow K_j |\Psi^{[n]}_{\rm QM}\rangle =0 \;, \forall j=1,\cdots , n\;,
\end{equation} 
which automatically implies that the only acceptable solutions to \eqref{papq} must have  each individual particle 
 evolving according to its own dynamical constraint (the model being interaction free for now).
 Observe also that as Eq.~(\ref{PPPP}) is symmetric under exchange of the particle indexes it
has no problem to act as constraint operator also in the case the particles are indistinguishable
(in particular it does not mix the complete symmetric part  ${\cal H}^{(n,{\bf S})}_{\bf E}$ of ${\cal H}^{\otimes n}_{\bf E}$
 with the complete anti-symmetric part  ${\cal H}^{(n,{\bf A})}_{\bf E}$). 
For instance, in the case of a Bosonic model governed by the positive energy KG equation~(\ref{KGEQ}),
the positivity requirement on the $K_j$ forces us to select (\ref{constr++}) (instead of (\ref{constr+})) as the proper single particles terms: accordingly for this model  the $n$-body constraint operator $K^{[n]}$  
can be identified with  \begin{eqnarray}\nonumber 
&&\!\!\!\!\!\!K^{[n]}_{\rm KG^+}:= \sum_{j=1}^n {(K_{\rm KG^+})}_j = 
\int d^4p_1 \cdots  \int d^4p_n  \; \sum_{j=1}^n   \\
&& \!\!\! \times  
\Big(\Theta(p^0_j)\; \overline{p_j} \cdot \underline{p_j} - m^2\Big)^2 
  |\overline{p}_{1};\cdots ; \overline{p}_{n}\rangle \langle \overline{p}_{1} ;\cdots ; \overline{p}_{n}|\;. \nonumber \\ \labell{constrenne} 
\end{eqnarray}
Similarly for Fermionic models we should identify the single-particle terms $K_j$ with the operator 
(\ref{constrDIR++}) instead of (\ref{constrDIR}). Accordingly in this case $n$-body constraint operator $K^{[n]}$
becomes 
\begin{eqnarray}
&& \!\!\!K^{[n]}_{\rm D}:= \sum_{j=1}^n {(K_{\rm D})}_j = \nonumber 
\sum_{\sigma_1}\int d^4p_1 \cdots\sum_{\sigma_n}  \int d^4p_n  \sum_{j=1}^n \\
&& \!\! \times 
 \lambda^2_{\sigma_j} (\overline{p}_j)  |\phi_{\sigma_1} (\overline{p}_1);\cdots ; \phi_{\sigma_n} (\overline{p}_n)\rangle \langle \phi_{\sigma_1} (\overline{p}_1) ;\cdots ; \phi_{\sigma_n} (\overline{p}_n)|\;. \nonumber \\
 \labell{constrenneD}
\end{eqnarray}

If the QM dynamical equations that rule the equation of motion of the particles are relativistically covariant  as in the cases of Eqs.~(\ref{constrenne}) and~(\ref{constrenneD}), 
then  the identities~(\ref{did}) and (\ref{didnot}) that in the single particle case allows us to connect the distributions 
of the observers $O$ and $O'$, translate into 
\begin{eqnarray}\label{didn} 
|\Psi^{[n]^\prime}_{\rm QM}\rangle = U^{\otimes n}_{\Lambda} |\Psi^{[n]}_{\rm QM}\rangle\;,  
\end{eqnarray} 
and
\begin{eqnarray} \nonumber 
&& \!\!\!\Psi^{[n]\prime}_{\rm QM} (\overline{x}_1,\sigma_1;\cdots; 
\overline{x}_n,\sigma_n ) = \!\!\! \!\!\!\sum_{\sigma_1,\cdots,\sigma_n} S_{\sigma'_1,\sigma_1}^{-1}(\Lambda)\cdots
S_{\sigma'_n,\sigma_n}^{-1}(\Lambda) \\\label{didnotnee} 
&&\qquad \times \Psi^{[n]}_{\rm QM} (\Lambda^{-1} \overline{x}_1,\sigma'_1;\cdots; 
\Lambda^{-1} \overline{x}_n,\sigma'_n )\;, 
\end{eqnarray} 
respectively. 
Notice also that setting $t_1=\cdots = t_n=t$ in the last one, invoking Eqs.~(\ref{chiave}) and (\ref{imoimpo}) 
we obtain 
 the connection between the spinor 3D wave-functions  of QM
$\Psi^{[n]}_{\rm QM} (\vec{x}_1,\sigma_1;\cdots; 
\vec{x}_n,\sigma_n|t)$ and $\Psi^{[n]\prime}_{\rm QM} (\vec{x}_1,\sigma_1;\cdots; 
\vec{x}_n,\sigma_n|t)$, that
$O$ and $O'$ would assign to the {\it same} quantum trajectory of the $n$ particles -- see Fig.~\ref{figurecorrespondence} for a schematic representation of this identity and App.~\ref{seclore} for a technical discussion. 
\subsection{QM/GEB correspondence in Fock space}\label{sec:NONINT}

In this section we generalize the 
 correspondence (\ref{las}) to the Fock space representation of GEB discussed in Sec.~\ref{FOCK}.
 At variance with what we did in Secs.~\ref{sec:primasec}  and~\ref{sec:MULTI}, here we start by first introducing the  constraint operator  (\ref{CONSTRAINT1}), and then
 show that  the associated solutions can be directly connected to those of QFT. 

 \subsubsection{Constraint operators} \label{sec:constraintop} 

To connect GEB to QFT, start by considering the case of a Bosonic model where each individual particle evolves according to the positive energy KG equation~(\ref{KGEQ}). As we have seen in the previous section, in the first quantization
formalism the constraint operator of the model is provided by~(\ref{constrenne}). 
When the total number of particles is fixed to $n$, the  first quantization version of the constraint operator (\ref{constrenne}) assigns a contribution 
$\Big(\Theta(p^0)\; \overline{p} \cdot \underline{p} - m^2\Big)^2$ to each particle with 
 4-momentum $\overline{p}$, specifically
 \begin{eqnarray} 
K^{[n]}_{\rm KG^+} |S(\overline p_1; \cdots ;\overline p_n)\> &=&
\sum_{j=1}^n \Big(\Theta(p_j^0)\; \overline{p}_j \cdot \underline{p}_j - m^2\Big)^2 \nonumber \\
&&\times 
|S(\overline p_1; \cdots ;\overline p_n)\>\;,  \label{defqn} 
 \end{eqnarray} 
 with $|S(\overline p_1; \cdots ;\overline p_n)\>$ the symmetric version of
 $|\overline p_1; \cdots ;\overline p_n\>$ (see App.~\ref{app:multi}).
 Exploiting the correspondence~(\ref{defBOSne}), 
 Eq.~(\ref{defqn})    can now be turned into its  second quantization form by identifying 
 $K^{[n]}_{\rm KG^+}$ with the Fock operator   
\begin{equation}
K^{(\rm Fock)}_{\rm KG^+} :=\int d^4p\; \left[ \Theta(p^0) \; {\overline{p}} \cdot {\underline{p}} -m^2\right]^2\; a^\dag_{\overline{p}} a_{\overline{p}} \;, 
 \labell{bconst} 
\end{equation} 
with $a^\dag_{\overline{p}}$, $a_{\overline{p}}$ the  creation and annihilation operators that 
obey the  canonical commutation rules (\ref{ccr1}).

In
the Fermionic case we proceed in similar fashion. In  this case, from (\ref{constrenneD}), we see we need to introduce 
a Fock number operator that counts how many particles of the system are in single particle states
described by the vectors $|\phi_{\sigma} (\overline{p})\rangle$. 
To construct such a term we 
introduce a new collection of annihilation operators 
\begin{eqnarray} \label{expressback} 
a_{\phi_\sigma(\overline{p})} &:=&  \sum_{\sigma'=1}^4 u^*_{\sigma',\sigma}(\vec{p}) \; 
a_{\overline{p},\sigma'}\;, 
\end{eqnarray} 
with 
 $u_{\sigma,\sigma'}(\overline{p})$  the unitary matrices that connects
 the vectors $|\phi_{\sigma} (\overline{p})\rangle$ with the vectors
 $|\overline{p},\sigma \rangle$ (see App.~\ref{appconstraintDIRAC}).
By construction they  fulfill the same anti-commutation rules of $a^\dag_{\overline{p},\sigma}$ and 
$a_{\overline{p},\sigma}$, i.e. 
\begin{equation}
\{ a_{\phi_\sigma(\overline{p})}, a^\dag_{\phi_\sigma'(\overline{p}')}\}  = \delta_{\sigma,\sigma'} \;\delta^{(4)}(\overline{p}- \overline{p}') \:, \
\{ a_{\phi_\sigma(\overline{p})}, a_{\phi_\sigma'(\overline{p}')}\}  = 0 \;, 
\end{equation} 
so that $a^\dag_{\phi_\sigma(\overline{p})}a_{\phi_\sigma(\overline{p})}$ is exactly the  number
operator we are looking for. Accordingly we can construct  the 
Fock counterpart of  (\ref{constrenneD}) by taking \begin{eqnarray} \label{fconstr} 
K^{(\rm Fock)}_{\rm D} &:=& \sum_{\sigma=1}^4 \int d^4p\; \lambda^2_\sigma(\overline{p}) \; a^\dag_{\phi_\sigma(\overline{p})}a_{\phi_\sigma(\overline{p})}\;,
\end{eqnarray}
which, via Eq.~(\ref{expressback}), can also be expressed as
\begin{equation} \label{fconstrnew} 
K^{(\rm Fock)}_{\rm D} = 
 \sum_{\sigma',\sigma''=1}^4  \int d^4p\;  D_{\sigma',\sigma''}(\overline{p}) \; a^\dag_{\overline{p},\sigma'}a_{\overline{p},\sigma''}(\overline{p})\;,
\end{equation}
with 
\begin{eqnarray} 
D_{\sigma',\sigma''}(\overline{p})&:=&  \sum_{\sigma=1}^4 u_{\sigma',\sigma}(\vec{p})
\lambda^2_\sigma(\overline{p}) u^*_{\sigma'',\sigma}(\overline{p})\nonumber \\
&=& \sum_{\sigma =1}^4 (\overline{\gamma}^\dag_{\sigma',\sigma} \cdot \underline{p})
(\overline{\gamma}_{\sigma,\sigma''} \cdot \underline{p}) + m^2 \delta_{\sigma',\sigma''} \nonumber \\
&&
- m (\overline{\gamma}^\dag_{\sigma',\sigma''} + \overline{\gamma}_{\sigma',\sigma''})\cdot \underline{p}\;,
\end{eqnarray} 
  where  we used~(\ref{defdife}) and (\ref{defdife1}). 
  
  \subsubsection{Connection with the QFT solutions}
Here we analyze   the solutions of the geometric constraint~(\ref{CONSTRAINT1}) 
that follow from the definitions~of $K^{(\rm Fock)}_{\rm KG^+}$ and $K^{(\rm Fock)}_{\rm D}$ given in the previous section, i.e.
 \begin{eqnarray}
 \int d^4p\; \left[ \Theta(p^0) {\overline{p}} \cdot {\underline{p}} -m^2\right]^2\; a^\dag_{\overline{p}} a_{\overline{p}} \; |\Psi_{\rm QM}\rangle=0 \;,\label{BOSONkg} 
 \end{eqnarray} 
 for the Bosonic model, and 
  \begin{eqnarray}
 \sum_{\sigma=1}^4 \int d^4p\; \lambda^2_\sigma(\overline{p}) \; a^\dag_{\phi_\sigma(\overline{p})}a_{\phi_\sigma(\overline{p})} \; |\Psi_{\rm QM}\rangle=0 \;,\label{DIRACkg} 
 \end{eqnarray} 
 for the Dirac one.

 It is clear that in both scenarios the no-event state, that in the
 theory is represented by 4D vacuum state
 $|\Psi_{\rm QM}\rangle =|0\rangle_4$ is an allowed solution. It corresponds to the trivial case of no particles (Bosons for \eqref{BOSONkg} or Fermions for \eqref{DIRACkg}) at all times. To
 discuss the other solutions in what follow we shall address first the
 Bosonic case that allows for some simplification due to the absence
 of spinor components.
\paragraph{Bosonic model:--} 
Express the vector 
 $|\Psi_{\rm QM}\rangle$ that appears on the r.h.s. of Eq.~(\ref{BOSONkg}) 
 as the one given in \eqref{superPP} (with no spin), namely
{ \begin{eqnarray}
&&\!\! \!\!\!\! \!\! \!\!  |\Psi_{\rm QM} \>:\stackrel{!}{=}  \label{psipospQMx} 
 \\\nonumber&&\!\! \!\! \!\! \sum_{n} 
  \frac{\alpha_n}{\sqrt{n!}} \left[ \prod_{j=1}^n \int d^4x_j 
 \;  a^\dag_{\overlinelz x_j}\right]|0\>_4 {\Psi}_{\rm QM}^{[n]}(\overline x_1,\cdots,\overline x_n
 ) \\\nonumber
&&\!\!\!\! \!\!  \!\!= \sum_{n}  \frac{\alpha_n}{\sqrt{n!}}  \left[ \prod_{j=1}^n\int d^4p_j \label{psipospQM} 
     \; a^\dag_{\overline p_j}\right] |0\>_4\tilde{\Psi}_{\rm QM}^{[n]}(\overline p_1; \cdots;\overline p_n) 
\end{eqnarray}
with 
 ${\Psi}_{\rm QM}^{[n]}(\overline x_1,\cdots,\overline x_n
 )$  and $\tilde{\Psi}_{\rm QM}^{[n]}(\overline p_1; \cdots;\overline p_n)$  connected via  4D Fourier transform:
\begin{align} 
&&\!\!\!\!\!\!\!\!\!\!{\Psi}_{\rm QM}^{[n]}(\overline x_1,\cdots
 ) =\left[ \prod_{j=1}^n\int \frac{d^4p_j}{4\pi^2}  e^{-i\overline{p}_j \cdot \underline{x}_j} \right]
  \:\tilde{\Psi}_{\rm QM}^{[n]}(\overline p_1,\cdots 
 )\;. \label{4FOURIER} 
 \end{align} }
 The functional dependence of the operator $K^{(\rm Fock)}_{\rm KG^+}$ upon the number operator
 $a^\dag_{\overline{p}}a_{\overline{p}}$ suggests to analyze Eq.~(\ref{BOSONkg}) in the 4-momentum representation~(\ref{psipospQMx}) instead of the position representation~(\ref{psipospQM}). Indeed 
 when acting on $a^\dag_{\overline p_1}\cdots a^\dag_{\overline p_n} |0\>_4$, $a^\dag_{\overline{p}}a_{\overline{p}}$  generates a multiplicative  factor $\sum_{j=1}^n  \delta(\overline p_j -\overline p)$ that allows us to 
 translate  the constraint (\ref{BOSONkg}) into a constraint on the momentum-representation (\ref{psipospene})  of the wavefunction as
 \begin{equation} \label{dfdsf} 
 \tilde{\Phi}_{\rm QM}^{[n]}(\overline p_1;\cdots;  \overline p_n
 )   \left( \sum_{j=1}^n \left(\Theta(p_j^0) {\overline{p}_j} \cdot {\underline{p}_j} -m^2\right)^2\right)  =0 \;. \end{equation} 
Such equation forces $\tilde{\Psi}_{\rm QM}^{[n]}(\overline p_1;\cdots;  \overline p_n
 )$ to have support only for values of  the  $\overline{p}_j$ momenta that satisfy the on-shell  condition ${\overline{p}_j} \cdot {\underline{p}_j} =m^2$ with $p_j^0\geq 0$. Specifically
 using 
\begin{eqnarray}\!\!\! \delta({\overline{p}} \cdot {\underline{p}}-m^2)\!\!\! &=&\!\!\! [\delta(p^0+E_p)+\delta(p^0-E_p)]/(2E_p)\;, \\
E_p&:=&+\sqrt{|{\vec p}|^{\:2}+m^2}\;, \label{defENP} 
\end{eqnarray}  
 we can express the most general solution of (\ref{dfdsf})  as 
 \begin{eqnarray} 
\!\! \tilde{\Psi}_{\rm QM}^{[n]}(\overline p_1;\cdots;  \overline p_n
 )  &=& \left(\Pi_{j=1}^n \delta\left( {\overline{p}_j} \cdot {\underline{p}_j} -m^2\right)
\right) \nonumber   \\ 
&& \times f^{[n]}(\overline p_1;\cdots;  \overline p_n
 )  \label{solgenKG} \\ 
&=&  \left(\Pi_{j=1}^n \delta(p_j^0-E_{p_j})
\right) \label{solcontrs}   \frac{f^{[n]}(\vec{p}_1;\cdots;  \vec{p}_n
 ) }{2 E_{p_1} \cdots 2 E_{p_n}},  \nonumber 
 \end{eqnarray}  
 \begin{eqnarray} 
\mbox{with } f^{[n]}(\vec p_1;\cdots;  \vec p_n
 )  := f^{[n]}(\overline p_1;\cdots;  \overline p_n
 ) \Big|_{p^0_j = E_{p_j}} 
 \end{eqnarray} 
 and $f^{[n]}(\overline p_1;\cdots;  \overline p_n
 ) $   an arbitrary function which nullifies for $p_j^0<0$ and that, in virtue of the
 implicit symmetry of (\ref{psipospQM}), 
   can always be forced to be  completely symmetric 
 under exchange of the indexes.
The  position representation ~(\ref{psipospQMx}) of the solution  can now be recovered replacing Eq.~(\ref{solgenKG}) into
 Eq.~(\ref{4FOURIER}), i.e. 
 \begin{eqnarray}
&&\!\!\!\!\!\!\!\!\!\!{\Psi}_{\rm QM}^{[n]}(\overline x_1,\cdots,\overline x_n
 )  =\Big[\prod_j\int \frac{d^3p_j}{(2\pi)^{3/2}} e^{i(\vec{p}_j \cdot \vec{x}_j-E_{p_j}{t}_j)}\Big] \nonumber \\ 
&&   \qquad \times
\frac{f^{[n]}(\vec{p}_1,\cdots,\vec{p}_n
 ) }{\sqrt{8 \pi}  E_{p_1} \cdots \sqrt{8 \pi} E_{p_n}} \;. \label{psipospQMnew}
 \end{eqnarray} 
To put these solutions  in correspondence with  the QFT solutions of the corresponding
 Bosonic KG field equation we observe that 
 in the Schr\"odinger picture, the general QFT solutions of a (positive-energy) Bosonic KG field equation writes as~\cite{peskin}
\begin{align} 
\!\!\!|\psi_{\rm QM}(t) \>=\sum_{n}  \frac{\beta_n}{\sqrt{n!}}  \Big[\prod_j\int {d^3x_j} \label{CONpsipospQM}  
\;   c^\dag_{\vec{x}_j}\Big]|0\>_3 {\Psi}_{\rm QM}^{[n]}(\vec{x}_1, \cdots,\vec{x}_n|t)  \;, 
\end{align}
where $\beta_n$ are normalized amplitude probabilities, $|0\rangle_3$ is the 3D vacuum state of the field (not to be confused with the 4-vacuum state $|0\rangle_4$ of GEB)  and the
$c^\dag_{\vec{x}}$'s are  Bosonic creation operators fulfilling the equal-time canonical commutation rules 
\begin{equation} 
 [c_{\vec{x}},c^{\dag}_{\vec{x}^{\;\prime}}]= \delta^{(3)}(\vec{x} - \vec{x}^{\;\prime})
, \
  [c_{\vec{x}},c_{\vec{x}^{\;\prime}}]=0\;.
\label{ccr1QM}
\end{equation} 
In the above expression ${\Psi}_{\rm QM}^{[n]}(\vec{x}_1, \cdots,\vec{x}_n|t)$ are (observer dependent) 3D+1 wave-functions that (under proper normalization conditions) define the joint probabilities of finding at time $t$, $n$ particles in $\vec{x}_1$, $\cdots$, $\vec{x}_n$: 
their temporal dependence is  fixed by the single-particle dispersion relation
defined in Eq.~(\ref{defENP}) and is computed in Eq.~(\ref{QFTIAO}). 
Our  goal is to show that  the GEB solutions~(\ref{psipospQM}) with $\tilde{\Psi}_{\rm QM}^{[n]}(\overline p_1;\cdots;  \overline p_n
 )$ as in Eq.~(\ref{solgenKG}) 
can be put in correspondence with~(\ref{CONpsipospQM}) 
by taking $\beta_n=\alpha_n$ and setting \begin{equation}  \label{definpsitilde} 
\tilde{\psi}_{\rm QM}^{[n]}(\vec{p}_1,\cdots, \vec{p}_n ) := \frac{f^{[n]}(\vec{p}_1,\cdots,\vec{p}_n
 )}{{\sqrt{8 \pi}  E_{p_1} \cdots \sqrt{8 \pi} E_{p_n}}} \;,
 \end{equation} 
 in Eq.~(\ref{QFTIAO}). 
 To verify this fact  notice that 
 for fixed value of $n\geq 1$ one can  invoke  Eqs.~(\ref{DEFVECTORSnbosNEW}) -- (\ref{defBOSne})
 to map the 4D wave-function ${\Psi}_{\rm QM}^{[n]}(\overline x_1,\cdots,\overline x_n
 )$ of Eq.~(\ref{psipospQMnew})  onto a QM 3D+1 wave-function of $n$ Bosonic particles via Eq.~(\ref{contratime}): 
this exactly reproduces the QFT solution $\Psi^{[n]}_{\rm QM} (\vec{x}_1,\cdots,
\vec{x}_n|t)$ of Eq.~(\ref{QFTIAO}) when we impose~(\ref{definpsitilde}). 

\paragraph{Fermionic model:--} 
Similar considerations apply to the Fermionic case. Here  the functional dependence of 
the constraint operator $K^{(\rm Fock)}_{\rm D}$ upon the number operator $a^\dag_{\phi_\sigma(\overline{p})}a_{\phi_\sigma(\overline{p})}$ suggests to expand the general solution \eqref{superPP}, namely
\begin{eqnarray}
|\Psi_{\rm QM} \>:\stackrel{!}{=}  \sum_{n}  \frac{\alpha_n}{\sqrt{n!}}\Big[\prod_j \sum_{\sigma_j=1}^4  \int d^4x_j\labell{psipospQMasddx} \; a^\dag_{\overline{x}_j, \sigma_j}\Big]
|0\>_4\\\nonumber \times  {{\Psi}}_{\rm QM}^{[n]}(\overline{x}_1, \sigma_1; \cdots;\overline{x}_n, \sigma_n)\;,
\end{eqnarray}
 in terms of the creation operators
$a^\dag_{\phi_\sigma(\overline{p})}$, i.e. 
\begin{eqnarray}|\Psi_{\rm QM} \>{=}  \sum_{n}  \frac{\alpha_n}{\sqrt{n!}} \Big[\prod_j\sum_{\sigma_j=1}^4  \int d^4p_j
\labell{psipospQMasd} \; a^\dag_{\phi_{\sigma_j}(\overline p_j)}\Big]|0\>_4 \\\nonumber\times\tilde{{\Psi}}_{\rm QM}^{[n]}({\phi_{\sigma_1}(\overline p_1)}; \cdots;{\phi_{\sigma_n}(\overline p_n)}) \;,
\end{eqnarray}
with the spinor wave-functions  ${{\Psi}}_{\rm QM}^{[n]}(\overline p_1, \sigma_1; \cdots;\overline p_n,\sigma_n)$
that are connected with those of the 4-position representation via the identity 
\begin{eqnarray} 
{\Psi}_{\rm QM}^{[n]}(\overline x_1,\sigma_1; \cdots; \overline x_n,\sigma_n
 ) =\Big[\prod_j\int \frac{d^4p_j}{4\pi^2}
e^{-i\overline{p}_j \cdot \underline{x}_j}
\nonumber\\
 \times 
\sum_{\sigma'_j=1}^4    u_{\sigma_j',\sigma_j}(\vec{p}_j) \Big]
\; \tilde{{\Psi}}_{\rm QM}^{[n]}({\phi_{\sigma'_1}(\overline p_1)}; \cdots;{\phi_{\sigma'_n}(\overline p_n)}) \;. \label{4FOURIERDIRACx} 
 \end{eqnarray} 
Replacing~(\ref{psipospQMasd}) into (\ref{DIRACkg}) we get 
  \begin{equation} \label{dfdsfsdsddd} 
 \tilde{{\Psi}}_{\rm QM}^{[n]}({\phi_{\sigma_1}(\overline p_1)}; \cdots;{\phi_{\sigma_n}(\overline p_n)})   \left( \sum_{j=1}^n \lambda^2_{\sigma_j}(\overline{p}_j) \right)  =0 \;, \end{equation} 
which due to the positivity of the terms $\lambda^2_{\sigma_j}(\overline{p}_j)$ has solutions of the form
\begin{eqnarray} 
  \tilde{{\Psi}}_{\rm QM}^{[n]}({\phi_{\sigma_1}(\overline p_1)}; \cdots;{\phi_{\sigma_n}(\overline p_n)})    \nonumber\\=\nonumber
 \left(\Pi_{j=1}^n \delta ( \lambda_{\sigma_j} (  \overline{p}_j) ) \right) \;  f^{[n]}(\overline p_1,\sigma_1;\cdots;  \overline p_n,\sigma_n
 )  \\ =  \left(\Pi_{j=1}^n \delta (  \overline{p}^0_j-E_{p_j}^{(\sigma_j)}) \right) 
\;  f^{[n]}(\vec{p}_1,\sigma_1;\cdots;  \vec{p}_n,\sigma_n
 ) \;, \nonumber 
\end{eqnarray} 
where   $E_{p}^{(\sigma)}=-E_p$ for $\sigma=1,3$ and $E_{p}^{(\sigma)}=E_p$ for $\sigma=2,4$ with $E_p=\sqrt{|\vec p|^2+m^2}$, see Eq.~(\ref{definEP}), and  
 where $f^{[n]}(\overline p_1,\sigma_1;\cdots;  \overline p_n,\sigma_n
 )$ are arbitrary functions that can always be assumed to completely anti-symmetric under particle indexes  exchange, 
 and  where finally 
 \begin{equation} 
  f^{[n]}(\vec{p}_1,\sigma_1;\cdots;  \vec{p}_n,\sigma_n
 )  : =f^{[n]}(\overline p_1,\sigma_1;\cdots;  \overline p_n,\sigma_n
 )\Big|_{p^0_j = E^{(\sigma_j)}_{p_j}} \;. 
 \end{equation} 
Substituting this into (\ref{4FOURIERDIRACx}) we hence obtain 
\begin{eqnarray} 
&&{\Psi}_{\rm QM}^{[n]}(\overline x_1,\sigma_1; \cdots; \overline x_n,\sigma_n
 )=\Big[\prod_j\int \tfrac{d^3p_j}{(2\pi)^{3/2}}\sum_{\sigma'_j=1}^4 
   \tfrac{u_{\sigma_j',\sigma_j}(\vec{p}_j)}{\sqrt{2\pi}}\nonumber\\
  && \times e^{i(\vec{p}_j \cdot \vec{x}_j-E^{(\sigma'_j)}_{p_j} t_j)} \Big]
 f^{[n]}(\vec{p}_1,\sigma'_1;\cdots;  \vec{p}_n,\sigma'_n)  \;.
 \label{4FOURIERDIRACxxx} 
 \end{eqnarray} 
 To establish a formal correspondence between (\ref{psipospQMasddx}) and  the solutions of QFT we observe that 
 the 3D+1 spinor wave-function of $n$ Fermionic particles that obey the Dirac equation is given by
 vectors of the form 
 \begin{eqnarray} 
|\psi_{\rm QM}(t) \>{=}\sum_{n}  \frac{\beta_n}{\sqrt{n!}}\Big[\prod_j  \sum_{\sigma_j=1}^4 \int {d^3x_j} \labell{CONpsipospQMD}   c^\dag_{\vec{x}_j,\sigma_j}\Big]
  |0\>_3\nonumber\\\times   
 \Psi^{[n]}_{{\rm QM}}(\vec{x}_1,\sigma_1;\cdots;\vec{x}_n,\sigma_n |t)  \;, \label{queq}
\end{eqnarray}
with creation operators that obey equal-time anti-commutation rules, i.e.  \begin{equation} 
 \{ c_{\vec{x},\sigma},c^{\dag}_{\vec{x}^{\;\prime},\sigma'}\} = \delta^{(3)}_{\sigma,\sigma'} \; \delta(\vec{x} - \vec{x}^{\;\prime})
, \
  \{ c_{\vec{x},\sigma},c_{\vec{x}^{\;\prime},\sigma'}\}=0\;.
\label{ccr1QMd}
\end{equation} 
and 3D+1 spinor wave-functions
$\Psi^{[n]}_{{\rm QM}}(\vec{x}_1,\sigma_1;\cdots;\vec{x}_n,\sigma_n
|t)$ defined in Eq.~(\ref{4FOURIERDIRACxxxN}).  Invoking once more
Eqs.~(\ref{DEFVECTORSnbosNEW}) -- (\ref{defBOSne}) we can hence
conclude that Eq.~(\ref{4FOURIERDIRACx}) corresponds to
(\ref{CONpsipospQMD}) by setting $\beta_n =\alpha_n$ and
$\phi^{[n]}(\vec{p}_1,\sigma'_1;\cdots; \vec{p}_n,\sigma'_n)$ of
Eq.~(\ref{4FOURIERDIRACxxxN}) equal to
$f^{[n]}(\vec{p}_1,\sigma'_1;\cdots; \vec{p}_n,\sigma'_n)$.   Note that here we also consider possible entanglement between
  different spinor components, whence the $n$ integrals in
  \eqref{queq}, which are not usually included in QFT treatments.
  Superpositions of a particle and an antiparticle are typically
  considered unphysical because one supposes that superselection rules
  will prevent them. However, such states have been proposed
  \cite{susskind,rudolph}, suggesting that superselection rules are
  never fundamental, but only practical limitations.  
\subsubsection{Lorentz transform}  
We conclude the section stressing that also in the Fock formalization of the model,
the constrained operators are Lorentz invariant quantities, allowing us to 
extend the identity (\ref{phitophi'nfock}) also to the elements  $|\Psi_{\rm QM}\rangle$ of ${\cal H}_{\rm QM}$, i.e. 
\begin{eqnarray} \label{phitophi'nfocknew}  
|\Psi'_{\rm QM} \rangle  = {\cal U}_{\Lambda} | \Psi_{\rm QM}\rangle \;, 
\end{eqnarray}
indicating that in GEB 
Lorentz transforms can be done entirely using unitary representations
of the Lorentz group as described above, according to Wigner's
prescription for symmetry transformations, entirely at the kinematic
level. This is clearly different to what happens in QFT where  we quantize ``on
shell'', namely, the quantization procedure contains the dynamics.
This implies for instance that the state $c^\dag_{\vec p}|0\>_3$ lives in a
${\mathcal L}^2({\mathbb R}^3)$ space of on-shell states, namely
states whose energy is $E_p$. In order to Lorentz transform such
state, one must first derive the new hyperboloid that satisfies
$E'^2_p-p'^2=m^2$ in the new frame and {\it then} quantize in the new
frame obtaining $c^\dag_{\vec p\:'}|0\>_3$ in the new frame (the
vacuum being Lorentz invariant).

\section{Conclusions}
\label{sec:conc} 
In conclusion we presented an alternative framework (GEB) for special
relativistic quantum mechanics. The full axiomatic structure of
quantum mechanics (e.g.~its statistical interpretation through the
Born rule) is applied covariantly.  The quantization is performed
axiomatically in GEB, constructing a Hilbert space for events, rather
than the customary QFT approach of quantizing the solutions of the
dynamical equations. The usual textbook relativistic QM and QFT are
obtained by conditioning over the temporal degrees of freedom of the
GEB event states.

We have not considered interactions here: as in relativistic QM and
QFT, interactions pose significant additional challenges
(understatement!) that will be tackled in future work. Other covariant
approaches that derive from Dirac forms \cite{diracforms} typically
work only for free particles (since the Hamiltonian ends up in the
boost generators): the ``no-interaction theorem''
\cite{orenstein,currie,leut}. Our approach might, instead, be able to
consider interactions, since we impose the dynamics only through a
constraint, which is a procedure known to bypass the no-interaction
theorem \cite{komar1,komar2,hor,hor1,ezrarov}.  Moreover, GEB does not
employ a quantization on the free-field dynamical equation solutions,
so it might perhaps be able to describe interacting fields without the
usual perturbative approach, if we will ever be able to devise
appropriate, solvable, constraint equations. GEB replies affirmatively
to a question raised by Kucha\u r \cite{kuchar} on whether the
constraint formalism is able to describe localized relativistic
particles (a completely different solution, based on the Newton-Wigner
mechanism, is in \cite{equivalencehoen}). Finally, it can treat
situations that do not admit a Hamiltonian formulation
\cite{rovelliham} (such as solutions to the KG equation without
positive-energy restriction or generic solutions of Einstein's field
equations \cite{peresgr,dirachamiltoniangr,wdw,pirani}), since, as
shown above, the constraint procedure does not require Hamiltonians to
describe the dynamics.

Of course, we do not claim that QFT is inadequate: the formulation
provided here is, as shown above, a (slight) extension of it and in
all situations considered in this paper an equivalent QFT description
exists ({\it mutatis mutandis}). It may perhaps be used to clarify some
longstanding problems, such as Haag's theorem \cite{teller} or
particle localization \cite{nw,teller} by recognizing that a localized
particle (that stays localized for a period of time) is not a physical
state (it does not satisfy the constraints), but it can be connected
to a kinematic state that can be used as an eigenstate of an
observable.

We believe that GEB opens new exciting avenues.

VG acknowledges feedback from A. Sagnotti, and financial support by MIUR (Ministero dell’Istruzione, dell’Universit\'{a} e della Ricerca) via project PRIN 2017 “Taming complexity via Quantum Strategies: a Hybrid Integrated Photonic approach” (QUSHIP) Id. 2017SRN- BRK, and via project PRO3 Quantum Pathfinder. SL acknowledges support from NSF, DOE, DARPA, AFOSR, and ARO.
LM acknowledges useful feedback from A. Bacchetta, A. Bisio, D.~E.
Bruschi, C.  Dappiaggi, G.  Carcassi, A. Smith, D. Wallace. This
material is based upon work supported by the U.S. Department of
Energy, Office of Science, National Quantum Information Science
Research Centers, Superconducting Quantum Materials and Systems Center
(SQMS) under contract number DE-AC02-07CH11359.

\appendix

\section{Notation and conventions used}\labell{s:notation}

\paragraph*{Physical Units:--} 
We use natural units setting $\hbar=1$ and $c=1$.
\\
\paragraph*{Spacetime coordinates:--} 
To represent 4D real vectors we use the notation  \begin{eqnarray}
\overline{a} := (a^0, a^1,a^2,a^3) = (a^0, \vec{a})\;,
\end{eqnarray} 
with $a^0$ the time-like component and $\vec{a}$ the associated {space}-like 3D vector
\begin{eqnarray}
\vec{a} := (a^1,a^2,a^3)\;.
\end{eqnarray} 
Greeks labels are employed to indicate the 4 components of $\overline{a}$, and 
roman labels to indicate the 3 components of $\vec{a}$; e.g.  
  $a^\mu$ with $\mu=\{ 0,1,2,3\}$ indicates the $\mu$-th term of $\overline{a}$, while 
  $a^i$ with 
   $i=\{ 1,2,3\}$ indicates the $i$-th term of~$\vec{a}$. 
Lower indexes 4D vectors are defined as
 \begin{eqnarray}
\underline{a} := (a_0, a_1,a_2,a_3) = (a^0, -\vec{a})\;,
\end{eqnarray} 
which are connected with their upper indexes counterpart via the transformations
\begin{eqnarray}
\underline{a} =\eta \; \overline{a}\;, \qquad \overline{a} =\eta \; \underline{a}\;, 
\end{eqnarray}  
 with $\eta$ the $4\times 4$ diagonal matrix 
\begin{align}
\eta :=\mbox{diag}(1,-1,-1,-1)
\labell{etamunu}\;,
\end{align}
defining the metric tensor of the theory whose elements are represented with the
 symbol
$\eta^{\mu\nu}=\eta_{\mu\nu}$. 
Recall next that given $\Lambda$ a $4\times4$ real matrix associated to a generic Lorentz transformation 
we have 
\begin{eqnarray} 
\Lambda^T \eta \Lambda = \eta\;, \label{property1} 
\end{eqnarray} 
from which it follows that 
 given the four-vectors $\overline{a}$ and $\overline{b}$ the
product
\begin{eqnarray}\label{defprodscalinv} 
\overline{a} \cdot \underline{b} := \overline{a} \eta \overline{b} = \sum_{\mu=0}^{3} a^\mu b_\mu\;,
\end{eqnarray} 
is an invariant quantity, i.e. $\overline{a} \cdot \underline{b}= \overline{a}' \cdot \underline{b}'$ with 
$\overline{a}' = \Lambda \overline{a}$ and $\overline{b}' = \Lambda \overline{b}$ 
(notice that the same term can also be computed as $\underline{a} \eta \underline{b}$ or as
 $\underline{a} \cdot \overline{b}$).
 
Special examples of 4-vectors are provided by the 4-position and 4-momentum
\begin{eqnarray} 
\overline{x} := (t, \vec{x}) \;, \qquad \overline{p} := (p^0, \vec{p}) \;,
\end{eqnarray} 
by the associated differential term $\partial^{\mu} = \frac{\partial}{\partial x_\mu}$, i.e. 
\begin{eqnarray}
\overline{\partial}= (\partial^0, \partial^1, \partial^2,\partial^3)=(\partial/\partial
{t},-\vec\nabla)\;.
\end{eqnarray} 
 Then,
\begin{align}
&X^\mu=\int d^4x\:x^\mu|\underline x\>\<\underline x|,\ 
P^\mu=\int d^4p\:p^\mu|\underline p\>\<\underline p| \Rightarrow
\nonumber\\\nonumber
&\<\underline x|\underline p\>=e^{-ip^\mu x_\mu}/({4\pi^2}),\mbox{
  namely }\\\nonumber
&\<x^0|p^0\>=\<{t}|E\>=\frac{e^{-iEt}}{\sqrt{2\pi}},\
  \<x^1|p^1\>=\<{x}|p_x\>=\frac{e^{+ip_xx}}{\sqrt{2\pi}},\cdots\\
&|\underline x\>\equiv c_x^\dag|0\>_4=
\int \frac{d^4p}{4\pi^2}e^{i\underline p\underline x}|p\>=\int
  \frac{d^4p}{4\pi^2}
e^{i\underline
  p\underline x}c^\dag_p|0\>_4\nonumber\\&
c_x^\dag=
\int
  \frac{d^4p}{4\pi^2}
e^{i\underline
  p\underline x}\:c^\dag_p,\ c_x=
\int
  \frac{d^4p}{4\pi^2}
e^{-i\underline
  p\underline x}\:c_p\;\nonumber\\&
c_{\vec x}^\dag=
\int
  \frac{d^3p}{(2\pi)^{3/2}}
e^{-i\vec 
  p\cdot\vec x}\:c^\dag_{\vec p},\ c_{\vec x}=
\int
  \frac{d^3p}{(2\pi)^{3/2}}
e^{i\vec
  p\cdot\vec x}\:c_{\vec p}\;
\labell{uffa}.
\end{align}
{{The $\gamma^\mu$ matrices are given by 
\begin{align}
\gamma^0=\left(\begin{matrix}\openone
    &0\cr0&-\openone\end{matrix}\right),\qquad 
\gamma^i=\left(\begin{matrix}0
    &\sigma_i\cr-\sigma_i&0\end{matrix}\right)
\labell{gammadef}\;,
\end{align}
with $\sigma^i$ being the Pauli operators. }}

\section{Connecting different reference frames} \label{S:reference} 

Consider first the simple case of a single spin-less event space. 
Let $O$ and $O'$  be two inertial observer whose 
coordinates are linked as in Eq.~(\ref{ddf}) of the main text with the $4\times 4$ matrix 
$\Lambda$ representing an element of the Lorentz group.
 Let  $\Phi(\overline{x})$ the wave-function of a state event  $S$ as described by $O$. To show that the observer $O'$ in his reference frame will describe it as the function $\Phi'(\overline{x})$ of (\ref{phitophi'amp}) 
 assume that $\Phi(\overline{x})$ gets its maximum value $\Phi_{\max}$ for $\overline{x} = \overline{x}_0$, i.e.
$\Phi_{\max} = \Phi(\overline{x}_0)$. 
The observer $O'$ will assign to such point the coordinate $\overline{x}_0' = \Lambda \overline{x}_0$ that represents the value at which $\Phi'(\overline{x})$ reaches its maximum, i.e.
\begin{equation} \nonumber 
\Phi'(\overline{x}'_0)=\Phi'( \Lambda \overline{x}_0)=\Phi_{\max} =  \Phi(\overline{x}_0)  \Longrightarrow  \Phi'( \Lambda \overline{x}_0)= \Phi(\overline{x}_0),
\end{equation} 
which leads exactly to~(\ref{phitophi'amp}).

\begin{figure}[hbt]
\begin{center}\epsfxsize=.9\hsize\leavevmode\epsffile{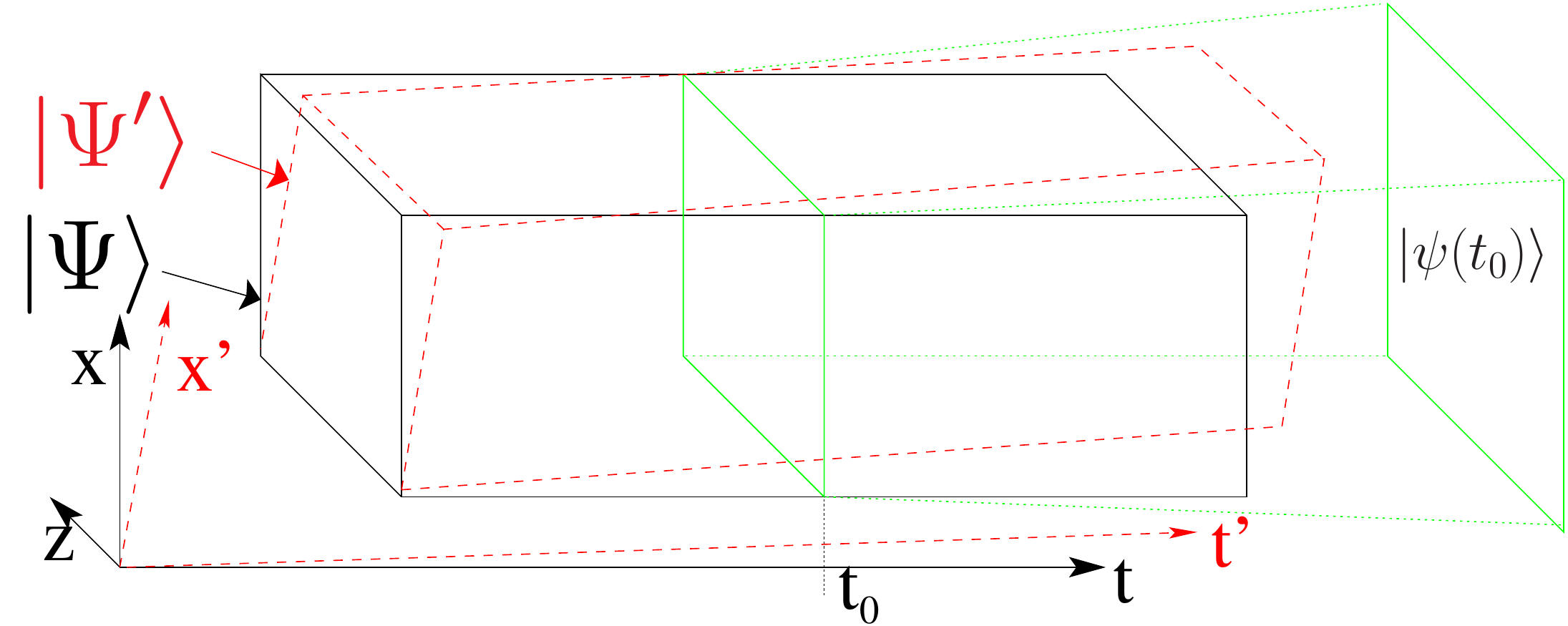}\end{center}
\vspace{-.5cm}
\caption{Figurative representation of the state $|\Psi\>$, its
  time-conditioning at time ${t}_0$ that gives the state
  $|\psi({t}_0)\>$, and its relativistic boost $|\Psi'\>$ to the
  reference $R'$ (dashed lines). The boosted reference $(x',t')$ is
  obtained through a hyperbolic transformation (Lorentz transform)
  from the $(x,{t})$ reference, pictorially represented with the
  dashed lines. The foliation in the $(x,{t})$ reference gives the
  usual (conditioned) state $|\psi({t}_0)\>$ of textbook quantum
  mechanics. A similar foliation in the $(x',t')$ reference (not
  pictured) is required for the quantum state at time $t'$ in the new
  reference.}
\labell{f:fig}\end{figure}

Let us now introduce the vectors $|\Phi\rangle$ and $|\Phi'\rangle$ of ${\cal H}_{\bf E}$ that $O$  and $O'$ will assign to the state $S$, i.e.
\begin{eqnarray} 
|\Phi\rangle &=& \int d^4{x} \; \Phi(\overline{x}) | \overline{x}\rangle\;, \\
|\Phi'\rangle &=& \int d^4{x}\; \Phi'(\overline{x}) | \overline{x}\rangle= \int d^4{x}\; \Phi(\Lambda^{-1} \overline{x}) | \overline{x}\rangle 
\nonumber \\
&=&\int d^4{x} \; \Phi( \overline{x}) | \Lambda \overline{x}\rangle \;,
\end{eqnarray} 
(Fig.~\ref{f:fig}).
By direct inspection one can be easily verify that these vectors fulfill the identity  (\ref{phitophi'}) of the main text 
by identifying the unitary transformation $U_\Lambda$ with the operator
\begin{eqnarray} \label{defunitarylambda} 
U_\Lambda &=& \int d^4{x} |\Lambda \overline{x}\rangle\langle \overline{x}|  = \int d^4{x} |\overline{x}\rangle\langle \Lambda^{-1} \overline{x}|\;, \\
U^\dag_\Lambda &=& \int d^4{x} | \overline{x}\rangle\langle \Lambda\overline{x}|  = \int d^4{x} |\Lambda^{-1} \overline{x}\rangle\langle \overline{x}|\;,  \label{defunitarylambdadag} 
\end{eqnarray} 
where the second identity in the first line follows by a simple chance of integration variables, while the second line is obtained by taking the adjoint of the first. 
Notice that $U_\Lambda$ and $U^\dag_\Lambda$ verify the conditions
\begin{eqnarray} 
U_\Lambda | \overline{x}\rangle &=& |\Lambda \overline{x}\rangle\;, \qquad \langle \overline{x} | U_\Lambda  = \langle \Lambda^{-1} \overline{x}| \;,\\
U_\Lambda^\dag | \overline{x}\rangle &=& |\Lambda^{-1} \overline{x}\rangle\;,  \qquad \langle \overline{x} | U^\dag_\Lambda  = \langle \Lambda \overline{x}|\;, 
\end{eqnarray} 
which represent the counterparts of~(\ref{ddf}) at the level of the generalized eigenstates of the position operator
$\overline{X}$. 
 Analogously for the generalized
eigenvectors of the momentum operator $\overline{P}$ we get
\begin{eqnarray}  \label{azionesup} 
U_\Lambda | \overline{p}\rangle &=& |\Lambda \overline{p}\rangle\;, \qquad \langle \overline{p} | U_\Lambda  = \langle \Lambda^{-1} \overline{p}| \;,\\
U_\Lambda^\dag | \overline{p}\rangle &=& |\Lambda^{-1} \overline{p}\rangle\;,  \qquad \langle \overline{p} | U^\dag_\Lambda  = \langle \Lambda \overline{p}|\;.
\end{eqnarray} 
The first for instance can be derived 
recalling Eq.~(\ref{defp}) and observing that 
\begin{equation} 
U_{\Lambda} |\overline p\> =\int \tfrac{d^4x}{4\pi^2}e^{-i\overline x\cdot \underline p}|\Lambda\overline x\>
=  \int \tfrac{d^4x}{4\pi^2}e^{-i(\Lambda^{-1} \overline x)\cdot \underline p}|\overline x\>
 = |\Lambda \overline p\>\;, 
\label{defpprimo} 
\end{equation} 
where in the last identity we exploit the invariance of the product~(\ref{defprodscalinv}) under Lorentz transform, i.e.
$\overline{a} \cdot \underline{b} = \overline{a}' \cdot \underline{b}'$, 
for $\overline{a}' = \Lambda \overline{a}$ and $\overline{b}' = \Lambda \overline{b}$.
\\

Consider next the expectation values of a generic operator $\Theta$ on  $S$. 
The observer $O$ will compute this as
\begin{eqnarray} 
\langle \Theta\rangle = \langle \Phi |\Theta | \Phi\rangle \;, 
\end{eqnarray} 
while $O'$ will see this as
\begin{eqnarray} 
\langle \Theta\rangle' = \langle \Phi' |\Theta | \Phi'\rangle =  \langle \Phi |  U^\dag_\Lambda \Theta U_\Lambda | \Phi\rangle\;,
\end{eqnarray} 
which of course needs not to be the same as $\langle \Theta\rangle$. Notice that we can also rewrite 
$\langle \Theta\rangle'=\langle \Phi |\Theta' | \Phi\rangle$ 
where now 
\begin{eqnarray} \label{transfr} 
\Theta'  = U^\dag_\Lambda \Theta U_\Lambda \;,
\end{eqnarray} 
is a sort of 4D ``Heisenberg-picture'' that allows us to transform the
operators instead of the states in moving from the reference frame of
$O$ to the one by $O'$.  In particular we shall say that $\Theta$ is
invariant under Lorentz transformations if $\Theta' = \Theta$ for all
choices of $\Lambda$, i.e.
\begin{eqnarray} \label{transfrinvariant} 
U^\dag_\Lambda \Theta U_\Lambda =\Theta \;,  \qquad \forall \Lambda\;, 
\end{eqnarray} 
while, given a collection  of operators  $A^0, A^1,A^2,A^3$ we shall call $\overline{A} = (A^0, A^1,A^2,A^3)$  a vectorial operator if 
\begin{eqnarray} \label{transfvectors} 
 U^\dag_\Lambda \; \overline{A} \; U_\Lambda  = \Lambda \overline{A}\;,  \qquad \forall \Lambda \;. 
\end{eqnarray} 
 From (\ref{property1}) it then follows that 
given $\overline{A}$ and $\overline{B}$ arbitrary vectorial operators the operator 
$\overline{A}\cdot \underline{B}$ is invariant. 
Important examples of vectorial operators are provided by the canonical operators $\overline{X}$ and  $\overline{P}$ of the theory, as anticipated in Eqs.~(\ref{heisenberg1}) and (\ref{heisenberg2}) of the main text. To see this explicit observe for instance that 
 from~(\ref{defP}) and (\ref{transfr})  we get 
 \begin{eqnarray} 
(X^\mu)'&=& U^\dag_\Lambda X^\mu U_\Lambda  = \int d^4{x} \; x^\mu   | \Lambda^{-1} \overline{x}\rangle\langle\Lambda^{-1}  \overline{x}| \nonumber \\
&=&
 \int d^4{x} \; (\Lambda \overline{x})^\mu   |  \overline{x}\rangle\langle  \overline{x}|=
( \Lambda \overline{X})^\mu \;,
\end{eqnarray} 
which leads to~(\ref{heisenberg1}).
We notice that the above expressions can be used to show that the  $U_\Lambda$'s admit as generators 
operators $M^{\mu, \nu}$ entering in  (\ref{poinc}):
for example, a $y$-axis rotation by an angle $\theta$
is generated by $U_{\Lambda}=e^{-i\theta M^{13}}$ with
$M^{13}=X^1P^3-X^3P^1$ so that
$U^\dag_{\Lambda} X^3U_{\Lambda}=X^3\cos\theta+X^1\sin\theta$
; a $x$-axis directed boost by
a rapidity $v$ is generated by $U_{\Lambda}=e^{-ivM^{01}}$ with
$M^{01}=X^1P^0-X^0P^1$ so that $U X^1U^\dag=X^1\cosh v+X^0\sinh v$
(the hyperbolic trigonometric functions appear because of the extra
minus sign in $[X^\mu,P^\nu]=-i\eta^{\mu\nu}$).
We stress also that (\ref{heisenberg1}) and  (\ref{heisenberg2}) are fully consistent with the setting of the problem.
In fact indicating with $\langle \overline{X}\rangle$ and $\langle \overline{X}\rangle'$  the mean position that $O$ and $O'$ assign to the same event  state we notice that they
are connected via the identity 
\begin{equation}  
\langle \overline{X}\rangle' := \langle \Phi | \overline{X'}  | \Phi\rangle = \langle \Phi | \Lambda  \overline{X}  | \Phi\rangle = \Lambda  \langle \Phi |  \overline{X}  | \Phi\rangle 
=  \Lambda \langle \overline{X}\rangle\;, \label{covarianceofx} 
\end{equation} 
which is exactly what you would aspect from Eq.~(\ref{ddf}).
Of course the same result can be obtained by working in the Schr\"{o}dinger picture: in this case in fact we get
\begin{eqnarray} 
&&\langle \overline{X}\rangle' = \langle \Phi' | \overline{X}  | \Phi'\rangle =\int d^4x\;   \overline x \; |\Phi'(\overline{x})|^2 
\\ 
&& \;\; = 
\int d^4x\;   \overline x \; |\Phi(\Lambda^{-1} \overline{x})|^2 = \int d^4x\;  \Lambda \overline x \; |\Phi(\overline{x})|^2 = \Lambda \langle \overline{X}\rangle \;, \nonumber 
\end{eqnarray} 
where in the third identity we used~(\ref{phitophi'amp}). 

\subsection{Spinors} \label{sec:spin} 
In the presence of spinorial degree of freedom, the 4D spinor wave-functions 
$\Phi'(\overline{x},\sigma')$ and $\Phi(\overline{x},\sigma)$ assigned by the observers $O'$ and $O'$,
will be connected as in Eq.~(\ref{phitophi'amp+spin}). 
This implies that Eqs.~(\ref{defunitarylambda}) and (\ref{defunitarylambdadag}) are replaced by 
\begin{eqnarray} \label{defunitarylambdasigma} 
U_\Lambda &=& \sum_{\sigma,\sigma'} S^{-1}_{\sigma,\sigma'}(\Lambda) \int d^4{x} |\Lambda \overline{x},\sigma'\rangle\langle \overline{x},\sigma| 
\nonumber \\
 &=& \sum_{\sigma,\sigma'} S^{-1}_{\sigma,\sigma'}(\Lambda) \int d^4{x} |\overline{x},\sigma' \rangle\langle \Lambda^{-1} \overline{x},\sigma|\;, \\
U^\dag_\Lambda &=&  \sum_{\sigma,\sigma'}  S_{\sigma,\sigma'}(\Lambda)  \int d^4{x} | \overline{x},\sigma'\rangle \langle \Lambda\overline{x},\sigma| \nonumber \\
&=& \sum_{\sigma,\sigma'}  S_{\sigma,\sigma'}(\Lambda)  \int d^4{x} |\Lambda^{-1}\overline{x},\sigma' \rangle\langle \overline{x},\sigma|\;, 
\end{eqnarray} 
so that 
\begin{eqnarray} 
U_\Lambda | \overline{x},\sigma\rangle &=&\sum_{\sigma'} S_{\sigma,\sigma'}^{-1}(\Lambda) |\Lambda \overline{x},\sigma'\rangle \;,\\
U_\Lambda^\dag | \overline{x},\sigma \rangle &=&\sum_{\sigma'} S_{\sigma,\sigma'}(\Lambda) |\Lambda^{-1} \overline{x},\sigma'\rangle\;,  \end{eqnarray}
and 
\begin{eqnarray} 
U_\Lambda | \overline{p},\sigma\rangle &=&\sum_{\sigma'} S_{\sigma,\sigma'}^{-1}(\Lambda) |\Lambda \overline{p},\sigma'\rangle \;,\\
U_\Lambda^\dag | \overline{p},\sigma \rangle &=&\sum_{\sigma'} S_{\sigma,\sigma'}(\Lambda) |\Lambda^{-1} \overline{p},\sigma'\rangle\;.\end{eqnarray}

\section{Multi-event tensor representation and Fock representation} \label{app:multi} 
Recall  
 that the projectors $\Pi^{(n,{\bf S})}$ and $\Pi^{(n,{\bf A})}$ associated with
the completely symmetric ${\cal H}^{(n,{\bf S})}_{\bf E}$ and the completely anti-symmetric 
${\cal H}^{(n,{\bf A})}_{\bf E}$ subspaces of ${\cal H}^{\otimes n}_{\bf E}$,
can be expressed as 
\begin{eqnarray} \label{defnv0} 
\Pi^{(n,{\bf S})} =\frac{1}{n!}  \sum_{\bf{p}}  V_{\bf{p}} \;, \quad  
\Pi^{(n,{\bf A})} = \frac{1}{n!}  \sum_{\bf{p}}  \mbox{sign}[{\bf{p}}] V_{\bf{p}} \;, 
\end{eqnarray} 
where the sums over $\bf{p}$ run on the set of permutations of $n$ elements, and $V_{\bf{p}}$ is  the
 unitary operator which represents ${\bf p}$  on ${\cal H}^{\otimes n}_{\rm E}$.

As mentioned in the main text the  $n$ event states of Bosonic QM/GEB
are by  vectors  $|\Phi^{[n]} \rangle$ of (\ref{DEFVECTORSn}) with 4D spinor wave-functions $\Phi^{[n]} (\overline x_1, \sigma_1;\cdots;\overline x_n,\sigma_n
 )$
 obeying  the symmetry condition~(\ref{symmBOS}) and 
  normalization condition
 \begin{equation} 
  \sum_{\sigma_1,\cdots,\sigma_n} \int d^4x_1 \cdots d^4x_n 
  |{\Phi}^{[n]} (\overline x_1, \sigma_1;\cdots;\overline x_n,\sigma_n
 )|^2 =1 \;. \label{normalizz} 
 \end{equation} 
Since these vectors  belong to the completely symmetric ${\cal H}^{(n,{\bf S})}_{\bf E}$  subspace of ${\cal H}^{\otimes n}_{\bf E}$ we have $|\Phi \rangle = \Pi^{(n,{\bf S})} |\Phi\rangle$  which exploiting~(\ref{defnv0}) allows one to 
equivalently rewrite Eq.~(\ref{DEFVECTORSn})  as
 \begin{eqnarray}  
 &&\!\!\!\!\!\!  |\Phi^{[n]}\rangle = \frac{1}{\sqrt{n!}} \sum_{\sigma_1,\cdots,\sigma_n} \int d^4x_1 \cdots d^4x_n 
 \label{DEFVECTORSnbosNEW} \\ \nonumber &&\qquad \times  {\Phi}^{[n]}(\overline x_1, \sigma_1;\cdots;\overline x_n,\sigma_n
 )
  |S(\overline x_1,\sigma_1; \cdots ;\overline x_n,\sigma_n)\>\;,  \end{eqnarray}
 with  
 \begin{eqnarray} \nonumber
  && \!\!\! \!\!\! |S(\overline x_1,\sigma_1; \cdots ;\overline x_n,\sigma_n)\>  := 
  \frac{1}{\sqrt{n!}}  \sum_{\bf{p}}  V_{\bf{p}} |\overline x_1,\sigma_1; \cdots ;\overline x_n,\sigma_n\> \\
   &&\quad \qquad  = \frac{1}{\sqrt{n!}}  \sum_{\bf{p}} |\overline x_{{\bf p}(1)},\sigma_{{\bf p}(1)}; \cdots ;
   \overline x_{{\bf p}(n)},\sigma_{{\bf p}(n)}\> \;  \label{simmetrixcp} 
 \end{eqnarray} 
  the completely symmetric counterpart of $|\overline x_1,\sigma_1; \cdots ;\overline x_n,\sigma_n\>$. 
 In the Fock space representation $|S(\overline x_1,\sigma_1; \cdots ;\overline x_n,\sigma_n)\>$ is  the vector (not
 $|\overline x_1,\sigma_1; \cdots ;\overline x_n,\sigma_n\>$) that is formally expressed as the application of sequences of the Bosonic creation operators $a^\dag_{\overlinelz x,\sigma}$'s to the 4-vacuum state, i.e. 
 \begin{equation}  \label{defBOSne} 
 a^\dag_{\overlinelz
  x_1, \sigma_1 }\cdots a^\dag_{\overlinelz x_n, \sigma_n}|0\>_4\Big|_{\rm BOS}  \equiv |S(\overline x_1,\sigma_1; \cdots ;\overline x_n,\sigma_n)\> \;, 
  \end{equation} 
  which replaced into~(\ref{DEFVECTORSnbosNEW}) leads to (\ref{psiposr}) 
 (to justify (\ref{defBOSne})  notice that due to the commutation rules  (\ref{ccr1}) the two family of states on the l.h.s. and the r.h.s. of the above equation have the same symmetry under permutation of indexes and the same scalar products).  

 Similar considerations apply for the Fermionic case where the 4D spinor wave-function appearing in (\ref{DEFVECTORSn}) 
 fulfill   the anti-symmetric relation~(\ref{symmFER}). Invoking hence the fact that they are elements of 
 the completely anti-symmetric ${\cal H}^{(n,{\bf A})}_{\bf E}$  subspace of ${\cal H}^{\otimes n}_{\bf E}$ we have now  $|\Phi \rangle = \Pi^{(n,{\bf A})} |\Phi\rangle$, which allows one to replace Eq.~(\ref{DEFVECTORSnbosNEW}) with 
  \begin{eqnarray} 
  &&\!\!\!\!\!\!  |\Phi^{[n]}\rangle =\frac{1}{\sqrt{n!}}  \sum_{\sigma_1,\cdots,\sigma_n} \int d^4x_1 \cdots d^4x_n 
   \label{DEFVECTORSnFERNEW}  \\\nonumber &&\qquad \times 
\Phi^{[n]}(\overline x_1, \sigma_1;\cdots;\overline x_n,\sigma_n
 ) 
  |A(\overline x_1,\sigma_1; \cdots ;\overline x_n,\sigma_n)\>\;,\end{eqnarray}
 with 
 \begin{eqnarray} \nonumber
  &&|A(\overline x_1,\sigma_1; \cdots ;\overline x_n,\sigma_n)\>   \\
  && \qquad :=\nonumber
  \frac{1}{\sqrt{n!}} \sum_{\bf{p}}  \mbox{sign}[{\bf{p}}] V_{\bf{p}} |\overline x_1,\sigma_1; \cdots ;\overline x_n,\sigma_n\> \\
&& \qquad  =  \frac{1}{\sqrt{n!}}  \sum_{\bf{p}} |\overline x_{{\bf p}(1)},\sigma_{{\bf p}(1)}; \cdots ;
   \overline x_{{\bf p}(n)},\sigma_{{\bf p}(n)}\> \; \nonumber
 \end{eqnarray}  the vector that is now identified by 
 sequences of Fermionic creation operator $a^\dag_{\overlinelz x,\sigma}$'s to the 4D-vacuum state, i.e. 
 \begin{equation} 
 a^\dag_{\overlinelz
  x_1, \sigma_1 }\cdots a^\dag_{\overlinelz x_n, \sigma_n}|0\>_4 \Big|_{\rm FER} \equiv  |A(\overline x_1,\sigma_1; \cdots ;\overline x_n,\sigma_n)\>\;,
\end{equation}  
leading  once more 
to (\ref{psiposr}).
 
\section{ More on the QM/GEB correspondence}\label{s:QMGEBcorrespondence} 
Here we analyze in detail the technical aspects of the QM/GEB correspondence introduced in Sec.~\ref{sec:CORR}.
Specifically we shall show 
 that  the vectors $|\Psi_{\rm{QM}}\>$ introduced in Eq.~(\ref{las}), 
 while not being elements of ${\cal H}_{\bf E}$, form a special subset ${\cal H}_{\bf QM}$ of the 
 distributions set ${\cal H}_{\bf E}^{+}$ of the theory, i.e. the rigged-extended version  of ${\cal H}_{\bf E}$ which we introduce when discussing the generalized position and momentum eigenvectors of GEB.

We have already commented the fact that  the normalization condition~(\ref{normaQM})  implies 
that  the $ |\Psi_{\rm QM}\>$'s of \eqref{las} have a divergent norm. This automatically excludes them from the Hilbert space 
${\cal H}_{\bf E}$. To prove that they are distributions, we 
need  to show that there exists a dense subset ${\cal D}$ of  ${\cal H}_{\bf E}$  formed by  (normalized) vectors $|\Phi\rangle$ such that the quantity $\langle \Psi_{\rm{QM}}|\Phi\rangle$ exists and is finite. 
To exhibit such subset let first introduce the spectral decomposition of the QM Hamiltonian $H$ which is ruling the dynamical evolution of the single-particle of the problem (i.e. the generator 
which is responsible for the  time evolution of the 3D wave-function $\Psi_{\rm{QM}}(\vec{x}|t)$).
We will consider explicitly the case where $H$ has a (possibly degenerate) continuous spectrum
but the analysis can be easily applied to the cases of discrete spectra (or even mixed discrete/continuous spectra).
Accordingly we write 
\begin{eqnarray}
H :=  \int d 
E \sum_k E | E,k\rangle\langle E,k | \;,
\end{eqnarray} 
with the discrete variable $k$ accounting for the degeneracy of the
$E$-energy level, and where $\{ |E,k\rangle\}_{E,k}$ are the
generalized  orthonormal eigenvectors that fulfill 
\begin{eqnarray}
\langle E',k' | 
 E,k\rangle = \delta_{k,k'} \delta(E-E') \;,
\end{eqnarray} 
with $\delta_{k,k'}$  the Kronecker delta symbol. Similarly to~\cite {paw,qtime}, we now adopt a spacetime foliation that separate the temporal coordinate of ${\cal H}_{\bf E}$  vs the spatial ones
  via a tensor product, writing 
  \begin{eqnarray}\label{foliation} 
  |\uline{x}\rangle= |t\rangle |\vec{x}\rangle\;,\end{eqnarray} (notice that while this choice breaks the covariance of the theory, this is not a problem as in our case we shall compute scalar products between vectors which are explicitly invariant quantities). We then expand 
 a generic normalized element of ${\cal H}_{\bf E}$ in the following form
  \begin{eqnarray}\label{DEC} 
  |\Phi\rangle =\int d E \sum_{n,k} c_{n,k}(E) |n\rangle |E,k\rangle \;, 
  \end{eqnarray} 
 where
 we introduced a discrete complete orthonormal set  $\{ |n\rangle\}_{n}$ for 
 the temporal axis while we adopted the generalized eigenstates $\{ |E,k\rangle\}_{E,k}$ of the  
 QM Hamiltonian $H$ to expand the spatial degree of freedom of the system. 
 In the above equation  
  $ c_{n,k}(E)$ are probability amplitudes   fulfilling the normalization condition
  \begin{eqnarray} \label{questa} 
\int dE  \sum_{n,k} |c_{n,k}(E)|^2=\langle \Phi| \Phi\rangle =1\;, 
 \end{eqnarray} 
Now we define ${\cal D}$ to be set of vectors of ${\cal H}_{\bf E}$  which
  admits a decomposition~(\ref{DEC}) with  coefficients  $c_{n,k}(E)$ that, besides~(\ref{questa}),  fulfill  also the extra constraint
  \begin{eqnarray}
 \sum_{n} \sqrt{\int dE  \sum_{k} |c_{n,k}(E) |^2 } < 
 \infty\;, \label{COND} 
  \end{eqnarray} 
  (to see that ${\cal D}$ is dense observe that such space contains all the vectors $|\Phi\rangle$
  with $c_{n,k}(E)\neq 0$ only for a finite set of values of $n$). 
  Expressing now $|\Psi_{\rm{QM}}\rangle$ of Eq.~(\ref{las}) in terms of the same spacetime foliation used in~(\ref{DEC}), i.e.
   \begin{equation}
 |\Psi_{\rm QM}\>=\int dt\int d^3x \Psi_{\rm{QM}}(\vec{x}|t) |t\rangle | \vec{x}\>=
\int dt |t\rangle | \psi(t) \>
\labell{lassemp}\;,
\end{equation}
with 
\begin{eqnarray} 
| \psi(t) \> = \int d^3x \Psi_{\rm{QM}}(\vec{x}|t)  | \vec{x}\>\;, \label{defPSI} 
\end{eqnarray} 
we notice that 
    \begin{eqnarray}
  \langle \Phi|\Psi_{\rm{QM}} \rangle &=& \int d t \int dE \sum_{n,k} c^*_{n,k}(E) \langle n |t\rangle  \langle E,k|\psi(t)\rangle \nonumber \\ 
  &=& \int d t \int dE \sum_{n,k} c^*_{n,k}(E) \langle n |t\rangle \alpha_k(E) e^{-i E t}  \nonumber \\
  &=&  \sqrt{2\pi} \int dE \sum_{n,k} c^*_{n}(E) \alpha_k(E) \langle n |\pi(E) \rangle, \label{HENCE}
   \end{eqnarray} 
   where in the second identity we introduced the probability amplitudes
   \begin{eqnarray} \alpha_k(E) e^{-i E t}:=  \langle E,k|\psi(t)\rangle\end{eqnarray}  of the state $|\psi(t)\rangle$ with $e^{-i E t}$ being their associated dynamical phase (remember that $\{ |E,k\rangle\}_{E,k}$ are eigenvectors of the system Hamiltonian), and where in the
   third identity we introduce the vectors
   \begin{eqnarray}
    |\pi(E)\rangle := \frac{1}{\sqrt{2\pi}} \int d t  e^{-i E t}|t\rangle\;.
   \end{eqnarray}  
   Observe that this last is a distribution for the temporal coordinate (indeed it is the Fourier transform of "position" coordinates), that fulfills the orthonormalization rule
   \begin{eqnarray} 
  \langle \pi(E')  |\pi(E)\rangle= \delta(E-E')\;. 
   \end{eqnarray} 
   As a matter of fact we can identify $|\pi(E)\rangle$ as a generalized eigenstate of the
   canonical momentum of the temporal position axis. Accordingly we can interpret $\langle \pi(E) |n\rangle$ as
   the momentum amplitude probability distribution of $|n\rangle$ evaluated at momentum $E$. 
   Remember next that $\{ |n\rangle\}_n$ is a basis that we can choose freely. We now take such basis as the
   orthonormal set of the spectrum of the Harmonic oscillator which allows us to explicitly compute the value of
    $\langle \pi(E) |n\rangle$ as
    \begin{eqnarray}
     \langle \pi(E) |n\rangle = \frac{\pi^{-1/4}}{\sqrt{2^n n!}} \; \exp[ - E^2/2] H_n(E) \;, 
    \end{eqnarray} 
    where for the sake of simplicity we are expressing here the function in renormalized units where all the physical constants are set equal to 1, and where $H_n(x)$ are the Hermite polynomials. Now the only 
    fundamental aspect of the problem here is that we can put an upper bound on such terms, independently of the choice of $n$ and $E$. In particular we can show that 
      \begin{eqnarray}
     \langle \pi(E) |n\rangle \leq \langle \pi(E)=0 |n=0\rangle  = \frac{1}{\pi^{1/4}}  \;.
    \end{eqnarray} 
    Hence invoking the Cauchy-Schwarz inequality, we can now bound the term (\ref{HENCE}) as follows:
        \begin{eqnarray}
  |\langle \Phi|\Psi_{\rm{QM}} \rangle| &\leq&  \sqrt{2\pi} \sum_{n}\int dE \sum_{k} \left| c^*_{n,k}(E) \alpha_k(E) \langle n |\pi(E) \rangle\right| \nonumber \\
  &\leq&  \frac{\sqrt{2\pi}}{\pi^{1/4}} \sum_{n}\int dE \sum_{k} \left| c^*_{n,k}(E) \alpha_k(E) \right| \nonumber \\
  &\leq&  \frac{\sqrt{2\pi}}{\pi^{1/4}} \sum_{n}\sqrt{ \int dE \sum_{k} | c^*_{n,k}(E)|^2 } \;, 
  \label{HENCE1}
   \end{eqnarray} 
   which is finite due to Eq.~(\ref{COND}). 

\section{Initial conditions} \label{secinitial}

The constraint Eq.~(\ref{CONSTRAINT1}) merely selects all possible distributions which are compatible with an assigned QM dynamical law. One can add extra constraints that enforce possibly observer dependent 
  ``initial'' (rather, boundary) conditions or better specify the system evolution.  
  For instance 
  we can identify  the element of ${\cal H}_{\bf QM}$ associated to the QM quantum trajectory 
  of a spin-less single-particle which at time $\tau$ as measured
for  the observer $O$, corresponds to a certain target 3D spinor wave-function  $\psi_{0}(\vec{x},\sigma)$, by 
looking for the $|\Psi_{{\rm QM}}\>$  fulfilling~(\ref{CONSTRAINT1}) which verifies the extra condition
 \begin{eqnarray} \Pi_{\tau} |\Psi_{{\rm QM}}\>= \sum_\sigma \int d^3x\;   \psi_{0}(\vec{x},\sigma) |\overline{x},\sigma \rangle \Big|_{t=\tau} 
\;, \label{CONSTRAINT2}\end{eqnarray}
 with 
  $\Pi_{\tau}  = \sum_\sigma \int d^4x \;  \delta(t-\tau) |\overline{x},\sigma\rangle\langle \overline{x},\sigma|$ 
 being a generalized projector on ${\cal H}_{\bf E}$.

%
\section{Constraint operator for the  KG model}\label{KGdynamics} 

In the absence of the energy constraint 
the general solution of the   KG equation 
    \begin{eqnarray}\label{KGEQunc} 
(\square +m^2)\Psi_{\rm KG}(\overline{x})=0 \;, \end{eqnarray} 
expressed in term of spacetime coordinates of an inertial observer~$O$ 
is given by the sum of two  independent contributions
\begin{eqnarray} 
\Psi_{\rm KG}(\overline{x}) &=& \Psi_{\rm KG}^{+}(\overline{x}) + \Psi_{\rm KG}^{-}(\overline{x}) \;, \label{soluzione}  \\ 
\Psi_{\rm KG}^{\pm}(\overline{x})  &:=& \int \frac{d^3 {p}}{(2\pi)^{3/2}}  e^{\mp i E_{p} t + i\vec{p}\cdot \vec{x}} \psi^{(\pm)}(\vec{p}) \;, 
\label{input} 
\end{eqnarray} 
with $E_{p}:=  \sqrt{|\vec{p}|^2 +m^2}$ and with the functions
$\psi^{(\pm)} (\vec{p})$  fixed by 
 imposing  boundary conditions.  Without introducing extra structure on the problem, Eq.~(\ref{soluzione}) is 
 not compatible with unitary evolutions predicted by QM since, as discussed below, the two parts can be seen as time evolutions according to two different Hamiltonian (e.g. given  $\Psi_{\rm KG,1}(\overline{x})$ and 
$\Psi_{\rm KG,2}(\overline{x})$ solutions of (\ref{KGEQunc}) we get 
$\int d^3 x  \Psi^*_{\rm KG,1}(\overline{x})  \Psi_{\rm KG,2}(\overline{x})$ is an explicit
function of $t$).
 Yet one still use Eq.~(\ref{las})  to associate to $\Psi_{\rm KG}(\overline{x})$ 
 a distribution $|\Psi_{\rm KG}\> $ of GEB and observe that the resulting vector 
 can be 
identified with the solutions of an eigenvalue equation~(\ref{CONSTRAINT1})  
 \begin{eqnarray}\label{idi} 
J_{\rm KG}|\Psi_{\rm KG} \> =0 \;,
\end{eqnarray} 
with
 constraint operator 
 \begin{equation}
J_{\rm KG} := \overline{P} \cdot \underline{P}-m^2 =
\int d^4p \; (\overline{p} \cdot \underline{p} - m^2) |\overline{p}\rangle \langle \overline{p}|\;, 
\labell{constr} 
\end{equation}
that  is  explicit Lorentz invariant. 
From the Eq.~(\ref{soluzione}) it follows that we can be written as $|\Psi_{\rm KG}\rangle$  the sum of two terms 
     \begin{eqnarray}
 |{\Psi}_{\rm KG} \> &=&\int d^4x\; \Psi_{\rm KG}(\overline
  x)\;|\overline x\>= \nonumber 
|\Psi_{\rm KG}^{+} \>  + |\Psi_{\rm KG}^{-} \> \;, 
\\
 |\Psi_{\rm KG}^{\pm} \> &:=& \label{defpsiqm} 
\int d^4x\; \Psi_{\rm KG}^{\pm}(\overline
  x)\;|\overline x\>\;, \end{eqnarray} 
  which also satisfy~(\ref{idi}), i.e. 
   \begin{eqnarray}\label{idipm} 
J_{\rm KG}|\Psi_{\rm KG}^{\pm} \> =0 \;.
\end{eqnarray} 
Selecting  the positive (negative) energy solutions of (\ref{soluzione}) corresponds   to
identifying $\Psi_{\rm KG}(\overline{x})$
with  just the component 
  $\Psi_{\rm KG}^{+}(\overline{x})$ (resp. $\Psi_{\rm KG}^{-}(\overline{x})$), i.e.
  to imposing  $\psi^{(-)} (\vec{p})=0$ (resp. $\psi^{(+)} (\vec{p})=0$) as boundary condition of the problem. 
 By construction, these special functions can  be seen as solutions of ordinary
Schr\"{o}dinger equations  with single-particle Hamiltonian $H:= \sqrt{ m^2-\nabla^2}$, i.e. 
\begin{eqnarray}\label{KGEQ11eq} 
 i \partial_t \Psi_{\rm KG}^{+}(\overline{x})&=&  H
 \Psi_{\rm KG}^{+}(\overline{x})\;, \end{eqnarray} 
(the same holds also for $\Psi_{\rm KG}^{-}(\overline{x})$, choosing $-H$ as Hamiltonian). 
Therefore, $\Psi_{\rm KG}^{+}(\overline{x})$  represents a proper unitary temporal evolution that preserves equal time, 3D scalar products.

A better insight on the properties of  
the distributions $|\Psi_{\rm KG}^{\pm} \>$  can be gained by 
rewriting~(\ref{defpsiqm}) as
\begin{eqnarray} |\Psi_{\rm KG}^{\pm} \> := 
\int d^4p \; \Psi_{\rm KG}^{\pm}(\overline
  p)\;|\overline p\>\;, 
\end{eqnarray} 
where $\Psi_{\rm KG}^{\pm}(\overline
  p)= \int \frac{d^4 {x}}{4\pi^2}  e^{ i \overline{x} \cdot \underline{p}} \;  \tilde{\Psi}_{\rm KG}^{\pm} (\overline{x})$ is the 4D  Fourier transform of  $\tilde{\Psi}_{\rm KG}^{\pm} (\overline{x})$
  which, by explicit computation, is given by 
 \begin{eqnarray}
\tilde{\Psi}_{\rm KG}^{\pm}(\overline{p}) &:=& \sqrt{2\pi} \delta(p^0\mp E_{p})  \psi^{(\pm)}(\vec{p})\;.
\end{eqnarray} 
Introducing the orthogonal projectors 
\begin{eqnarray} 
\Pi^{+} &:=& \int d^4 p \; \Theta( p^0) |\overline{p}\rangle\langle \overline{p}|\;,\\
\Pi^{-} &:=& \openone_{\rm E} - \Pi^{+} = \int d^4 p \; \Theta( -p^0) |\overline{p}\rangle\langle \overline{p}|\;,
\end{eqnarray} 
that identify the positive/negative energy subspaces of ${\cal H}{\rm E}$,
we note that they admit $|\Psi_{\rm KG}^{\pm} \>$ as eigenvectors that solve the identities 
\begin{eqnarray} \Pi^{+} |\Psi_{\rm KG}^{+} \> = |\Psi_{\rm KG}^{+} \> \;,\quad 
 \Pi^{-} |\Psi_{\rm KG}^{-} \> = |\Psi_{\rm KG}^{-} \>\;.
\end{eqnarray} 
or equivalently 
\begin{eqnarray} 
 \Pi^{-} |\Psi_{\rm KG}^{+} \> =0 \;,\quad 
 \Pi^{+} |\Psi_{\rm KG}^{-} \> = 0 \;,
\end{eqnarray} 
Thanks to (\ref{idipm}) this allows us to 
 uniquely identify $|\Psi^{+}_{\rm KG} \>$  as the special vectors
  which are in the intersection of the kernels of $J_{\rm KG}$ and $\Pi^{-}$, i.e. 
\begin{eqnarray}
J_{\rm KG^+}|\Psi_{\rm KG}^{+} \> =0 \;,
\end{eqnarray} 
with the new constraint operator
\begin{eqnarray} \nonumber 
J_{\rm KG^+}&:=&J_{\rm KG} \Pi^+ - m^2\Pi^-=\Pi^+ J_{\rm KG}- m^2 \Pi^- \\
&=&  \int d^4p \;\left[  \Theta(p^0) \; \overline{p} \cdot \underline{p} - m^2\right]  |\overline{p}\rangle \langle \overline{p}|\;,
\end{eqnarray} 
(note the $-m^2\Pi^-$ term!)
reported in Eq.~(\ref{constr+}) of the main text.
Notice that such a term is explicitly self-adjoint ($J^\dag_{\rm KG^+}=
J_{\rm KG^+}$), but not positive semidefinite (indeed its generalized eigenvalues $\Theta(p^0) \; \overline{p} \cdot \underline{p} - m^2$ can take any real values for proper choices of the 4-momentum $\overline{p}$). Similarly the negative energy terms can be
  uniquely identified by writing
  $J_{\rm KG^-}|\Psi_{\rm KG}^{-} \> =0$ with
\begin{align} J_{\rm KG^-}&:=& J_{\rm KG} \Pi^- -m^2 \Pi^+=\Pi^- J_{\rm KG}- m^2\Pi^+ \\
  &=& \int d^4p \; \left[   \Theta(-p^0)\;  \overline{p} \cdot \underline{p} - m^2\right]|\overline{p}\rangle \langle \overline{p}|\;.
\end{align}

Consider next what happens when we introduce a new observer $O'$  sitting in a reference frame $R'$ whose 4D coordinates $\overline{x}'$  are connected with those of $O$ via the mapping~(\ref{ddf}).
Due to the explicit covariant structure of~(\ref{KGEQ}), in the new reference frame the
general solution $\Psi_{\rm KG}(\overline{x})$ is replaced by the new function 
\begin{eqnarray}\label{trasnfpsi'} 
\Psi'_{\rm KG}(\overline{x}) = \Psi_{\rm KG}(\Lambda^{-1} \overline{x}) \;, 
\end{eqnarray} 
which corresponds to the identity~(\ref{didnot})  which at the level of the correspondence~(\ref{las}), 
leads to Eq.~(\ref{did})  of the main text. 
To verify that the same holds for the positive (negative) solutions as well, 
the important observation  is that  these functions do not mix under Lorentz transformations.
Specifically  one can verify that  $\Psi'_{\rm KG}(\overline{x})$ 
still maintain the same
structure of~(\ref{soluzione}),  \begin{eqnarray} 
\Psi'_{\rm KG}(\overline{x}) &=& \Psi'^{+}_{\rm KG}(\overline{x}) + \Psi'^{-}_{\rm KG}(\overline{x}) \;, \label{soluzione'}  
\end{eqnarray} 
with  new positive and negative energy terms
\begin{equation} 
\Psi'^{\pm}_{\rm KG}(\overline{x})  = \int \frac{d^3 {p}}{(2\pi)^{3/2}}  e^{\mp i E_{p} t + i\vec{p}\cdot \vec{x}} \psi^{'(\pm)}(\vec{p}) \;, 
\end{equation} 
 that are associated with those of $O$ via the  same coordinate change of
(\ref{trasnfpsi'}), i.e.  
\begin{eqnarray} \Psi'^{\pm}_{\rm KG}(\overline{x}) = \Psi^{\pm}_{\rm KG}(\Lambda^{-1} \overline{x})\;.\label{covpos} 
\end{eqnarray} 
For instance assuming $\Lambda$  to represent a boost along the 
$x$ direction (i.e. $t'=  \gamma(t-vx)$,  $x' = \gamma (x-vt)$, $y'=y$, and $z'=z$)
we get 
${\psi}^{'(\pm)} (\vec{p})  =  \psi^{(\pm)} (\gamma(p^1\pm v E_{p}), p^2,p^3))  \tfrac{\gamma (E_{p} 
\pm vp^1)}{E_{p}}$ which  shows the independence of ${\Psi}'^{+}_{\rm KG} (\overline{x})$ 
(${\Psi}'^{-}_{\rm KG} (\overline{x})$) from ${\Psi}^{-}_{\rm KG}(\overline{x})$ (resp. ${\Psi}^{+}_{\rm QM}(\overline{x})$). 
 An important consequence of the property (\ref{covpos}) is that it implies  that
  we can drop the negative energy terms in 
 Eq.~(\ref{soluzione}) without affecting the Lorentz invariance of Eq.~(\ref{KGEQ})
 hence ensuring that also the non explicitly covariant Eq.~(\ref{KGEQ11eq}) yields
Lorentz  covariant solutions (this is exactly what we need to
show that 
 Eq.~(\ref{did}) also applies in the special case  where we focus on the positive (negative) solutions 
of the KG equation~(\ref{KGEQ})).

 We now briefly comment on the physical
  significance of the negative energy solutions of the Klein-Gordon
  equation. Remember that the wave equation
  $(\square-m^2) f({t},\vec r)=0$ has solutions with spacetime
  dependence $f=g(\vec r-\vec v {t})+h(\vec r+\vec v {t})$, with
  $\vec v$ the propagation velocity (both signs of the velocity must
  appear in the general solution as the wave equation contains only
  $v^2$).  One can expand $g$ and $h$ in terms of plane waves
  $e^{i\vec k\cdot(\vec r\pm\vec v {t})}\equiv e^{i(\vec k\cdot\vec
    r-\omega {t})}$, where the frequency
  $\omega\equiv\mp\vec k\cdot\vec v$ can be positive or negative
  depending on the propagation {\it direction} of the wave with
  respect to the wave vector $\vec k$. With an appropriate choice of
  sign in the definition of $\omega$, one can consider a
  negative-frequency wave as an {\it advanced} solution to the wave
  equation and a positive-frequency wave as a {\it retarded}
  solution, since these solutions can be
  obtained from one another by time reversal.  Usually the advanced
  solution is discarded (set to zero) appealing to some vague notion
  of causality, e.g.~\cite{griffiths}, but more careful analyses
  \cite{jackson,einstein1909} interpret the retarded solutions as a
  prediction based on past boundary conditions and the advanced
  solutions as a retrodiction based on future boundary conditions.
  Then the choice of which frequency sign to choose (or even a
  combination of the two \cite{einstein1909}) is dictated purely by
  the available boundary conditions. Clearly, past boundary conditions
  are more useful in general. One can discard the negative frequency
  solutions by imposing, in addition to the Klein-Gordon equation of
  motion, an additional {\it physical} condition of positive-energy
  (as was done in the main text).

In closing we comment on the ``negative probability densities'' that
historically have plagued the acceptance of the Klein-Gordon equation
(notoriously, it was discovered, but then discarded, by Schr\"odinger
\cite{schweber,bjorken1}). This problem ensues from the observation
that, if one defines a four-current for the Klein-Gordon wave-function
$\psi_1$ as
$j^\mu=\psi_1^*\partial^\mu\psi_1-\psi_1\partial^\mu\psi_1^*$, it does
satisfy a conservation equation $\partial_\mu j^\mu=0$, but the
density $j^0$ (representing a putative probability density) is not
positive definite (and should be interpreted as a charge density).  It
is not such $j^0$ that should take the role of a probability density
of the particle position {\it at a certain time}, but rather
$|\psi_1(\overline x)|^2$ that is the probability density of finding a
particle-detection event at space{\it time} position
$\overline x=({t},\vec x)$: a joint probability for both the position
and for time, rather than a conditioned probability for the position,
given the time. As such, $|\psi_1(\overline x)|^2$ is a {\it scalar}
quantity, not the temporal component of a 4-current, and needs not
satisfy any current conservation. Moreover, it is obviously always
positive definite. In contrast, in the case of the Dirac field, one
can build also a (conserved) probability current (see below).

\section{Constraint operator for the Dirac model} \label{appconstraintDIRAC} 
The Dirac equation for the spinor wave-function $\Psi_{{\rm QM}}(\vec{x},\sigma|t)$ of single particle is a collection of the four differential equations reported in Eq.~(\ref{DIRAC}). 
By taking the 4D Fourier transform we can turn them into the equivalent form
\begin{eqnarray} 
\sum_{\sigma=1}^4  (\overline{\gamma}_{\sigma',\sigma} \cdot \underline{p} - m\; \delta_{\sigma',\sigma})
 \tilde{\Psi}_{{\rm QM}}(\overline{p},\sigma)  =0 \;,  \label{eqDIRCA} 
\end{eqnarray} 
with 
\begin{eqnarray} 
\tilde{\Psi}_{{\rm QM}}(\overline{p},\sigma) &=& \int \frac{d^4 x }{4\pi^2} \exp[i \overline{p} \cdot \underline{x}] \Psi_{{\rm QM}}(\vec{x},\sigma|t)\;.
\end{eqnarray} 
Contracting the index $\sigma'$ of~(\ref{eqDIRCA}) with the matrix elements of the invertible matrix $\gamma^0$, we can further modify Eq.~(\ref{DIRAC}) into the identity 
\begin{eqnarray} 
\sum_{\sigma=1}^4  M_{\sigma',\sigma}(\overline{p})  \tilde{\Psi}_{{\rm QM}}(\overline{p},\sigma)  =0 \;, \label{eqDIRCA1} 
\end{eqnarray}
where  
\begin{eqnarray} 
M_{\sigma',\sigma''}(\overline{p}) := \sum_{\sigma=1}^4 \gamma^0_{\sigma',\sigma} (\overline{\gamma}_{\sigma,\sigma''} \cdot \underline{p} - m\; \delta_{\sigma,\sigma''}) 
 \;, \label{defdife} 
\end{eqnarray} 
are elements of the self-adjoint (yet not positive)  $4\times 4$ matrix
 \begin{eqnarray} 
 M(\overline{p})  
&:=&\left(\begin{matrix}(p^0 - m)\openone
    &-\vec{\sigma}\cdot \vec{p}  \cr -\vec{\sigma}\cdot \vec{p}&(p^0 + m)\openone\end{matrix}\right),\label{equJ} 
\end{eqnarray} 
with eigenvalues 
\begin{eqnarray} \label{deflambdaDnew} 
\lambda_\sigma(\overline{p}) &:=&     p^0 - E^{(\sigma)}_{p} \;, \end{eqnarray}
where given $E_{p}=  \sqrt{|\vec{p}|^2 +m^2}$ we introduced the quantities
\begin{eqnarray} 
 E_{p}^{(\sigma)}  &: =& \left\{ 
 \begin{array}{ll}  
 - E_p & \mbox{for $\sigma =1,3\;,$}    \\ \\ 
 E_p & \mbox{for $\sigma =2,4\;,$} 
 \end{array}\right. \label{definEP} 
\end{eqnarray}  
%
 Equation~(\ref{eqDIRCA}) can hence be interpreted as an eigenvector equation which, for any assigned $\overline{p}$,  selects eigenvectors of  
$M(\overline{p})$ which are associated with 
null eigenvalues ($\lambda_\sigma(\overline{p})=0$).
More precisely casting $M(\overline{p})$ in diagonal form  
\begin{eqnarray}
M_{\sigma',\sigma''}(\overline{p}) = \sum_{\sigma=1}^4 u_{\sigma',\sigma}(\vec{p})\;   \lambda_{\sigma}(\overline{p}) \; 
u^*_{\sigma'',\sigma} (\vec{p})\;, \label{defdife1} 
\end{eqnarray} 
with $u_{\sigma,\sigma'}(\vec{p})$ the elements of a $4\times 4$ unitary  matrix (see the end of the section for explicit expressions),  it follows that the most generic solution of Eq.~(\ref{eqDIRCA}) can writes  as \begin{eqnarray}
\tilde{\Psi}_{{\rm QM}}(\overline{p},\sigma) = \sum_{\sigma'=1}^4\; \delta(p^0 - E_{p}^{(\sigma')}) \alpha_{\sigma'}(\vec{p}) \; 
u_{\sigma,\sigma'}(\vec{p}) \;, 
\end{eqnarray} 
with $\alpha_{\sigma'}(\vec{p})$  arbitrary functions, i.e. 
\begin{eqnarray}
\Psi_{{\rm QM}}(\vec{x},\sigma|t)&=& 
\int \frac{d^3 p }{(2\pi)^{\frac{3}{2}}}e^{i \vec{p} \cdot \vec{x}} 
 \sum_{\sigma'=1}^4\;  
 \frac{u_{\sigma,\sigma'}(\vec{p})}{\sqrt{2\pi}} \nonumber \\
&& \times e^{- i E_{p}^{(\sigma')} t}\; {\alpha_{\sigma'}(\vec{p})} 
\;, \label{singleDIRAC} 
\end{eqnarray} 
at the level of the 3D+1 spinor wave-function.
\\

Expressed as in Eq.~(\ref{eqDIRCA})  it is  easy to verify that, at the level of the GEB distribution
$|\Psi_{{\rm QM}}\rangle = \sum_{\sigma=1}^4\int d^4 x \;  \Psi_{{\rm QM}}(\vec{x},\sigma|t) |\overline{x}, \sigma\rangle$, the Dirac equation~(\ref{DIRAC}) corresponds to the identity 
$J_{\rm D} |\Psi_{{\rm QM}}\rangle  = 0$ 
with $J_{\rm D}$ as in Eq.~(\ref{constrDIR}).
Indeed, to show this, we need the fact that  thanks to~(\ref{defp}) 
$\tilde{\Psi}_{{\rm QM}}(\overline{p},\sigma)$ provides the 
4D-momentum spinor wave-functions expansion of  $|\Psi_{{\rm QM}}\rangle$, i.e. 
$|\Psi_{{\rm QM}}\rangle = \sum_{\sigma=1}^4\int d^4 p \;  \tilde{\Psi}_{{\rm QM}}(\overline{p},\sigma) |\overline{p}, \sigma\rangle$. 
As mentioned in the main text the operator $J_{\rm D}$ is not a self-adjoint: this is a direct consequence of the fact that for all $i=1,2,3$ the matrices $\gamma^i$ are anti-Hermitian (indeed $(\gamma^i)^\dag = - \gamma^i=\gamma_i$), while $\gamma^0$ is Hermitian, so that  
$J_{\rm D}^\dag= \sum_{\mu=1}^4 (\gamma^{\mu})^\dag P_\mu - m = \sum_{\mu=1}^4 \gamma_{\mu} P_\mu - m \neq J_{\rm D}$.
Notice  however that exploiting the fact that $\gamma^0\gamma^0 =\openone$, and 
$\gamma^0\gamma^i =\left(\begin{matrix}0
    &\sigma_i\cr\sigma_i&0\end{matrix}\right)$, 
we can write 
\begin{eqnarray}
J_{\rm D} =\gamma^0  J_{\rm D}^{(H)}\;, 
\end{eqnarray} 
where given  the matrix elements $M_{\sigma',\sigma''}(\overline{p})$ of Eq.~(\ref{defdife})
$J_{\rm D}^{(H)}$ is the self-adjoint operator
 \begin{eqnarray} 
J_{\rm D}^{(H)}:=  \sum_{\sigma',\sigma''} \int d^4p \; M_{\sigma',\sigma''}(\overline{p})  |\overline{p},\sigma' \rangle \langle \overline{p},\sigma''| 
\label{equJ1} \;.
\end{eqnarray} 
Equation~(\ref{constrDIR++}) finally follows by using Eq.~(\ref{defdife1}) observing that  the vectors   
\begin{eqnarray} \label{defphiu} 
|\phi_{\sigma} (\overline{p})\rangle := \sum_{\sigma'=1}^4 u_{\sigma',\sigma}(\vec{p}) | \overline{p},\sigma^\prime\rangle  \;,
\end{eqnarray}  
obey generalized orthonormal conditions~(\ref{neworthonw}) thanks to the  unitary properties of the matrix elements $u_{\sigma',\sigma}(\vec{p})$: indeed with this choice Eq.~(\ref{equJ1}) becomes 
\begin{eqnarray} 
J_{\rm D}^{(H)} = \sum_{\sigma=1}^4 \int d^4 p \;  \lambda_\sigma(\overline{p}) \; |\phi_\sigma(\overline{p})\rangle
\langle \phi_\sigma(\overline{p})| \;, 
\end{eqnarray} 
and hence 
\begin{eqnarray} 
K_{\rm D} &=& J_{\rm D}^\dag J_{\rm D} = \left( J^{(H)}_{\rm D}\right)^2 
\nonumber \\
&=&
\sum_{\sigma=1}^4 \int d^4p \; \lambda^2_\sigma (\overline{p}) \;  |\phi_\sigma (\overline{p})\rangle \langle \phi_\sigma (\overline{p})|\;. 
\end{eqnarray} 

We conclude reporting 
explicit expressions for the $|\phi_\sigma(\overline{p})\rangle$: 
\begin{widetext} 
\begin{eqnarray} \label{primaprima} 
|\phi_1(\overline{p})\rangle &:=&\frac{1}{\sqrt{2}} 
\Big( \sqrt{1- \frac{m}{E_{p}}}|\overline{p}, s_1(\hat{n}) \rangle - \frac{|\vec{p}|}{\sqrt{1- \frac{m}{E_{p}}}} 
|\overline{p}, s_2(\hat{n}) \rangle \Big) \;, \\
|\phi_2(\overline{p})\rangle &:=& 
\frac{1}{\sqrt{2}} 
\Big( \sqrt{1+ \frac{m}{E_{p}}}|\overline{p}, s_1(\hat{n}) \rangle + \frac{|\vec{p}|}{\sqrt{1+ \frac{m}{E_{p}}}} 
|\overline{p}, s_2(\hat{n}) \rangle \Big)
\;, \\
|\phi_3(\overline{p})\rangle &:=&\frac{1}{\sqrt{2}} 
\Big(\sqrt{1- \frac{m}{E_{p}}}|\overline{p}, s_3(\hat{n}) \rangle + \frac{|\vec{p}|}{\sqrt{1- \frac{m}{E_{p}}}} 
|\overline{p}, s_4(\hat{n}) \rangle \Big)\;, \\
|\phi_4(\overline{p})\rangle &:=& 
\frac{1}{\sqrt{2}} 
\Big( \sqrt{1+ \frac{m}{E_{p}}}|\overline{p}, s_3(\hat{n}) \rangle - \frac{|\vec{p}|}{\sqrt{1+ \frac{m}{E_{p}}}} 
|\overline{p}, s_4(\hat{n}) \rangle \Big)
\;, 
\end{eqnarray} 
where for $\hat{n} : = \vec{p}/|\vec{p}|$, $|\overline{p}, s_\sigma(\hat{n}) \rangle$ are the orthonormal vectors 
\begin{eqnarray} 
|\overline{p}, s_1(\hat{n}) \rangle &:=& \frac{1}{\sqrt{2}} \left(\sqrt{1+n^3} \; |\overline{p}, 1\rangle +
 \frac{n^1 + i n^2}{ \sqrt{1+n^3}} \; |\overline{p}, 2\rangle\right)\;,\\
 |\overline{p}, s_2(\hat{n}) \rangle &:=&   \frac{1}{\sqrt{2}} \left(\sqrt{1+n^3}\; |\overline{p}, 3\rangle +
 \frac{n^1 + i n^2}{ \sqrt{1+n^3}} \; |\overline{p}, 4\rangle\right)\;,\\
 |\overline{p}, s_3(\hat{n}) \rangle &:=&  \frac{1}{\sqrt{2}} \left(\sqrt{1-n^3} \;|\overline{p}, 1\rangle -
 \frac{n^1 + i n^2}{ \sqrt{1-n^3}} \; |\overline{p}, 2\rangle\right) \;,\\
 |\overline{p}, s_4(\hat{n}) \rangle &:=&  \frac{1}{\sqrt{2}} \left(\sqrt{1-n^3} \;|\overline{p}, 3\rangle -
 \frac{n^1 + i n^2}{ \sqrt{1-n^3}} \; |\overline{p}, 4\rangle\right) \;.\label{lastlast} 
\end{eqnarray} 
Observe that via Eq.~(\ref{defphiu}) these identities implicitly define   the matrix
 elements $u_{\sigma,\sigma'}(\vec{p})$: for instance we get 
 \begin{eqnarray} u_{1,1}(\vec{p})&=& \frac{1}{{2}} \sqrt{1- \frac{m}{E_{p}}}\sqrt{1+n^3}\;, \qquad 
 u_{2,1}(\vec{p})=\frac{1}{{2}} \sqrt{1- \frac{m}{E_{p}}} \frac{n^1 + i n^2}{ \sqrt{1+n^3}}\;,\\
 u_{3,1}(\vec{p})&=&- \frac{1}{{2}}  \frac{|\vec{p}|}{\sqrt{1- \frac{m}{E_{p}}}}\sqrt{1+n^3}\;, \qquad 
 \cdots \end{eqnarray} 
\end{widetext}

If one appropriately normalizes the state, one can recover, just as
for Bosons, a {\it scalar} probability density for each spinor
component, since the Dirac equation implies the Klein-Gordon one: 
$(\gamma^\mu p_\mu+m)(\gamma^\nu p_\nu-m)=(p^\mu
p_\mu-m^2)\openone_4$. In addition, one can, as usual, also introduce a
conserved 4-current
$j^\mu\equiv\Psi^\dag(\overline x)\gamma^0\gamma^\mu\Psi(\overline
x)=\bar\Psi(\overline x)\gamma^\mu\Psi(\overline x)$, where $\Psi$
is the column vector of the {\it conditioned} spinors in the position
representation, namely the column of position-representation
amplitudes.  Since the zeroth component
$j^0=\Psi^\dag(\overline x)\Psi(\overline x)$ is positive definite,
it can be given a probabilistic interpretation as the {\it
  conditional} probability density of finding a particle at position
$\vec x$, given that time is $t$, where $\overline x=({t},\vec x)$.

\section{Extra observations on the multi-event QM/GEB correspondence} \label{app:uniq1} 
This section is dedicated to making explicit some technical aspects of the  QM/GEB correspondence
in multi-event scenario discussed in Sec.~\ref{sec:MULTI}.
We start by showing that  the vector~(\ref{lasgen})  is uniquely defined; then we 
verify  that 
Eq.~(\ref{didnotnee}) gives the right prescription to compute the evolution of a QM 3D+1 spinor wave-function under Lorentz transformations.

\subsection{Uniqueness of the the multi-event QM/GEB correspondence}\label{app:uniq} 
Here we prove that the vector~(\ref{lasgen}) is uniquely defined.

To begin with recall that, for all $t$, the  QM spinor 3D wave-function $\Psi^{[n]}_{\rm QM}(\vec{x}_1,\sigma_1; \cdots; \vec{x}_n, \sigma_n|t)$ of $n$ particles can be expressed 
as 
\begin{equation} \label{defpsi3d} 
\Psi^{[n]}_{\rm QM}(\vec{x}_1,\sigma_1; \cdots; \vec{x}_n, \sigma_n|t) = \langle \vec{x}_1,\sigma_1; \cdots; \vec{x}_n, \sigma_n|
\psi^{[n]}_{\rm QM} (t) \rangle\;, 
\end{equation} 
 where 
\begin{eqnarray}
&&|\psi^{[n]}_{\rm QM} (t) \rangle = \sum_{\sigma_1,\cdots, \sigma_n} \int d^3 {x}_1 \cdots \int d^3 {x}_n\\ \nonumber 
&& \quad  \quad \Psi^{[n]}_{\rm QM}(\vec{x}_1,\sigma_1; \cdots; \vec{x}_n, \sigma_n|t) |\vec{x}_1,\sigma_1; \cdots; \vec{x}_n, \sigma_n\rangle \;, 
\end{eqnarray} 
is the associated wave-vector, and $|\vec{x}_j, \sigma_j\rangle$ the generalized 3D position eigenvectors of the $j$-th particle.
Recalling then that we are dealing with non-interacting systems, 
we can now write  
\begin{eqnarray} \label{lba} 
|\psi^{[n]}_{\rm QM} (t) \rangle = U_1(t) \otimes\cdots \otimes U_n(t)  |\psi^{[n]}_{\rm QM} (0) \rangle\;, 
\end{eqnarray} 
where for $j=1,\cdots, n$, $U_j(t)$ stands for the QM unitary transformation that rules the free evolution of the
$j$-th particle. Equation~(\ref{lba}) leads to Eq.~(\ref{chiave}) by 
expressing $|\psi^{[n]}_{\rm QM} (0) \rangle$ in terms of an arbitrary local basis for the $n$ particles, i.e. 
\begin{equation} \label{lba1} 
|\psi^{[n]}_{\rm QM} (0) \rangle = \sum_{\vec{\ell}} \alpha_{\vec{\ell}}\;  |\psi^{(\ell_1)} _{\rm QM} \rangle\otimes 
\cdots \otimes |\psi^{(\ell_n)} _{\rm QM}  \rangle \;, \end{equation}
and using the identities 
\begin{eqnarray} 
 \Psi^{(\ell_j)} _{\rm QM} (\vec{x}_j, \sigma_j|t)&:=&  \langle \vec{x}_j,\sigma_j|
 U_j(t) |\psi^{(\ell_j)} _{\rm QM} \rangle \;.
\end{eqnarray}

Replacing this into (\ref{imoimpo}) and 
(\ref{lasgen}) we finally arrive to 
\begin{eqnarray}\nonumber 
&& |\Psi^{[n]}_{\rm QM}\>=\sum_{\sigma_1,\cdots, \sigma_2} 
  \int d^4x_1\cdots  \int d^4x_n  |\overline{x}_1,\sigma_1; \cdots;  \overline{x}_n, \sigma_n\rangle  \\ \nonumber 
  &&  \; \times   \langle \vec{x}_1,\sigma_1; \cdots ; \vec{x}_n,\sigma_n|
 U_1(t_1) \otimes \cdots \otimes U_n(t_n)
|\psi^{[n]}_{\rm QM} (0) \rangle\nonumber , 
\labell{lasgenidentity}
\end{eqnarray}
which explicitly  shows that $|\Psi_{\rm QM}\>$ carries no functional dependence upon the specific choice
of the local basis $\{ |\psi^{(\ell)} _{\rm QM} \rangle\}_\ell$ used in~(\ref{lba1}).

 As an application of the above identities we report here the special cases of particles obeying 
 to the positive energy KG equation and the Dirac equation.
 For the KG equation, setting 
 $\tilde{\psi}_{\rm QM}^{[n]}(\vec{p}_1, \cdots,\vec{p}_n) := \sum_{\vec{\ell}} \alpha_{\vec{\ell}}
{\psi}^{(\ell_1)}(\vec{p}_1)  \cdots {\psi}^{(\ell_2)}(\vec{p}_n)$ and replacing (\ref{input}) into   Eq.~(\ref{chiave}) we get
 \begin{widetext} 
  \begin{eqnarray} 
 {\Psi}_{\rm QM}^{[n]}(\vec{x}_1, \cdots,\vec{x}_n|t) 
&=& \int \frac{d^3p_1}{(2\pi)^{3/2}} \cdots
  \frac{d^3p_n}{(2\pi)^{3/2}}  e^{i(\vec{p}_1 \cdot \vec{x}_1 + \cdots +\vec{p}_n \cdot \vec{x}_n)}  
 e^{-i (E_{p_1} + \cdots + E_{p_n})t} \; 
\tilde{\psi}_{\rm QM}^{[n]}(\vec{p}_1, \cdots,\vec{p}_n)\;, 
 \label{QFTIAO} 
\end{eqnarray} 
with associated 4D GEB spinor wave-function 
\begin{eqnarray} 
{\Psi}_{\rm QM}^{[n]}(\overline{x}_1, \cdots,\overline{x}_n) 
&=& \int \frac{d^3p_1}{(2\pi)^{3/2}} \cdots
  \frac{d^3p_n}{(2\pi)^{3/2}}  e^{i(\vec{p}_1 \cdot \vec{x}_1 + \cdots +\vec{p}_n \cdot \vec{x}_n)}  
 e^{-i (E_{p_1} t_1+ \cdots + E_{p_n}t_n)} \; 
\tilde{\psi}_{\rm QM}^{[n]}(\vec{p}_1, \cdots,\vec{p}_n)\;.
 \label{QFTIAOm1} 
\end{eqnarray} 
Similarly for the Dirac equation
from Eq.~(\ref{singleDIRAC}), setting 
\begin{eqnarray} \phi^{[n]}(\vec{p}_1,\sigma'_1;\cdots;  \vec{p}_n,\sigma'_n)
:= \sum_{\vec{\ell}} \alpha_{\vec{\ell}}\;  {\alpha_{\sigma'_1}(\vec{p}_1)} \cdots {\alpha_{\sigma'_n}(\vec{p}_n)}\;,\end{eqnarray} we get 
\begin{eqnarray} 
\Psi^{[n]}_{{\rm QM}}(\vec{x}_1,\sigma_1;\cdots;\vec{x}_n,\sigma_n |t)&=&\int \frac{d^3p_1}{(2\pi)^{3/2}} \cdots
  \frac{d^3p_n}{(2\pi)^{3/2}}  \sum_{\sigma'_1=1}^4    \frac{u_{\sigma_1',\sigma_1}(\vec{p}_1)}{\sqrt{2\pi}}\cdots
 \sum_{\sigma'_n=1}^4 \frac{ u_{\sigma_n',\sigma_n}(\vec{p}_n) }{\sqrt{2\pi}} 
  \; e^{i(\vec{p}_1 \cdot \vec{x}_1 + \cdots +\vec{p}_n \cdot \vec{x}_n)} \nonumber \\ 
&&   \qquad \times
e^{ -i (E^{(\sigma'_1)}_{p_1} + \cdots + E^{(\sigma'_n)}_{p_n}) t}  \; 
 \phi^{[n]}(\vec{p}_1,\sigma'_1;\cdots;  \vec{p}_n,\sigma'_n)  \;.
 \label{4FOURIERDIRACxxxN} 
 \end{eqnarray} 
 which at the level of GEB corresponds to 
\begin{eqnarray} 
\Psi^{[n]}_{{\rm QM}}(\overline{x}_1,\sigma_1;\cdots;\overline{x}_n,\sigma_n )&=&\int \frac{d^3p_1}{(2\pi)^{3/2}} \cdots
  \frac{d^3p_n}{(2\pi)^{3/2}}  \sum_{\sigma'_1=1}^4    \frac{u_{\sigma_1',\sigma_1}(\vec{p}_1)}{\sqrt{2\pi}}\cdots
 \sum_{\sigma'_n=1}^4 \frac{ u_{\sigma_n',\sigma_n}(\vec{p}_n) }{\sqrt{2\pi}} 
  \; e^{i(\vec{p}_1 \cdot \vec{x}_1 + \cdots +\vec{p}_n \cdot \vec{x}_n)} \nonumber \\ 
&&   \qquad \times
e^{ -i (E^{(\sigma'_1)}_{p_1}t_1 + \cdots + E^{(\sigma'_n)}_{p_n} t_n)}  \; 
 \phi^{[n]}(\vec{p}_1,\sigma'_1;\cdots;  \vec{p}_n,\sigma'_n)  \;.
 \label{4FOURIERDIRACxxxNg} 
 \end{eqnarray}

\end{widetext}

\subsection{Lorentz transformations}\label{seclore}

Equation~(\ref{defpsi3d}) represents the (time-dependent) 3D+1 spinor wave-function that an
observer $O$ would assign to describe the state of the $n$ particles on his
reference frame $R$. Assuming the dynamical evolution is relativistic consistent 
(e.g. the particles obey KG or Dirac dynamical equations), 
 we are now interested in determining the spinor 3D wave-function a second observer $O'$ sitting in the reference frame $R'$ with 4D coordinates $\overline{x}'$ that are linked with those of $R$ as in Eq.~(\ref{ddf}) will assign to such a state. 
Since particles are independent (i.e. no interactions are present in the model),
 this can be done using the decomposition~(\ref{chiave}) for $\Psi^{[n]}_{\rm QM} (\vec{x}_1,\sigma_1;\cdots; 
\vec{x}_n,\sigma_n|t$   and applying the single-particle transformation~(\ref{didnot}) to each individual term
$\Psi^{(\ell)}_{\rm QM}(\vec{x},\sigma| t)$.  Accordingly we can write 
 \begin{eqnarray} \nonumber 
&&\!\!\!\!\!\! \Psi^{[n]\prime}_{\rm QM} (\vec{x}_1,\sigma_1;\cdots; 
\vec{x}_n,\sigma_n|t) =\!\!\! \!\!\! \sum_{\sigma_1,\cdots,\sigma_n} S_{\sigma'_1,\sigma_1}^{-1}(\Lambda)\cdots
S_{\sigma'_n,\sigma_n}^{-1}(\Lambda) \\ 
&&\!\!\!\!\!\!\times \sum_{\vec{\ell}} \alpha_{\vec{\ell}} \; 
\Psi^{(\ell_1)}_{\rm QM}(\Lambda^{-1}\overline{x}_1,\sigma'_1)|_{t_1=t} \cdots \Psi^{(\ell_n)}_{\rm QM}(\Lambda^{-1}\overline{x}_n,\sigma'_n)|_{t_n=t}\;, \nonumber \\
\label{didnotnee1}
\end{eqnarray} 
whose r.h.s. exactly matches with the one of Eq.~(\ref{didnotnee}) of the main text. 
It is important to stress that  while not immediately evident from the resulting expression $\Psi^{[n]\prime}_{\rm QM} (\vec{x}_1,\sigma_1;\cdots; 
\vec{x}_n,\sigma_n|t)$ does not depends upon the specific choice of the local decomposition used in~(\ref{lba1}). One easy way to verify this is e.g. to use the fact that 
(\ref{lasgen}) does not depends on such a choice (see previous section) and the fact that 
thanks to Eq.~(\ref{didnotnee}) we can write 
\begin{eqnarray} 
&&\Psi^{[n]\prime}_{\rm QM} (\vec{x}_1,\sigma_1;\cdots; 
\vec{x}_n,\sigma_n|t)\\ \nonumber 
&& \qquad = \langle \overline{x}_1, \sigma_1 ;\cdots;
\overline{x}_n, \sigma_n|\Psi^{[n]\prime}_{\rm QM}\rangle\Big|_{t_1=\cdots=t_n=t}\\
&&\qquad= \langle \overline{x}_1, \sigma_1 ;\cdots;
\overline{x}_n, \sigma_n|U^{\otimes n}_{\Lambda} |\Psi^{[n]}_{\rm QM}\rangle\Big|_{t_1=\cdots=t_n=t} \;, \nonumber \end{eqnarray} 
where we made use of the invertion formula~(\ref{contratime}) and  of Eq.~(\ref{didn}).

\end{document}